\documentclass{aa}
\usepackage{times}
\usepackage{graphicx}
\usepackage{natbib}
\usepackage{mathptm}

\begin{document}

\title{Red-sequence galaxies with young stars and dust: \\
	The cluster Abell 901/902 seen with COMBO-17}

\author{C. Wolf\inst{1} \and M. E. Gray\inst{2} \and K. Meisenheimer\inst{3} }

\institute{ Department of Physics, Denys Wilkinson Bldg.,
            University of Oxford, Keble Road, Oxford, OX1 3RH, U.K. 
       \and School of Physics and Astronomy,
            University of Nottingham, Nottingham, NG7 2RD, U.K.  
       \and Max-Planck-Institut f\"ur Astronomie, K\"onigstuhl 17,
            D-69117 Heidelberg, Germany 
}

\date{Received 07 June 2005 / Accepted 03 August 2005}

\abstract{
We report the discovery of a rich component of dusty star-forming galaxies
contaminating the red-sequence in the supercluster system comprising Abell 
901a, 901b and A902 at redshift $\sim 0.17$. These galaxies do not fit into
the colour-density relation, because their preferred habitat is different
from that of regular red-sequence galaxies, which are typically dust-poor,
old and passively evolving. The dusty red galaxies prefer the medium-density
outskirts of clusters while being rare in both the low-density field and the
high-density cluster cores. 
This new result is based on the information content in the medium-band 
photometry of the COMBO-17 survey. The photo-z accuracy of the $\sim 800$ 
brightest cluster galaxies is $<0.01$ and of the order of the velocity 
dispersion of the cluster. This enables us to select a rich and clean 
cluster sample, in which we can trace age-sensitive and dust-sensitive
spectral features independently with the detailed medium-band SED data.
We find the red colour of the dusty galaxies to be a result of dust 
extinction combined with relatively old stellar ages.
We speculate that the dusty red galaxies could either be a product of 
minor mergers between established old red cluster galaxies with infalling
blue field galaxies, or mark a period in the internal transformation of 
blue field galaxies into red cluster galaxies, which is triggered by the
environmental influences experienced during cluster infall.  
}
\titlerunning{Dusty red-sequence galaxies in Abell 901/902}
\authorrunning{Wolf et al.}
\maketitle

\section{Introduction}

It has been known for decades that massive galaxies in clusters tend to have 
spheroid-dominated morphologies \citep{Dr80,Dr97} and a paucity of recent star
formation \citep{BO84,L02}. Dressler's morphology-density relation can be seen 
as a result of an environmental influence on galaxy properties, whereby a higher 
density of galaxies supresses star formation and leads to pressure-supported
morphologies via some still controversial physical mechanism. Alternatively, the 
galaxies in denser
regions could just be exhibiting a more mature state because they were the 
earliest to collapse and evolved most quickly. Both concepts could explain the
observation that the percentage of red old galaxies increases with the local
galaxy density, while that of blue young galaxies decreases correspondingly.

So far, it appears that clusters conform with rather universal trends of
evolution in colour-magnitude and colour-density relations \citep{McI05}, if
only a simple bimodal distinction of red vs. blue galaxies is considered, which 
is most simply translated into non-star-forming vs. star-forming galaxies.
However, from ISOCAM observations Coia et al. (2005) have concluded that some 
of the red-sequence galaxies in the cluster Cl 0024+1654 are undergoing bursts
of star formation while their red colour is due to their very dusty nature.

In this paper, we present striking evidence that the red-sequence of the
clusters Abell 901/902 is highly contaminated by galaxies which are not just old
and passively evolving. Instead we find a large population of dusty star-forming
galaxies in the red sequence, which do not fit into the simple colour-density
relation. These dusty red cluster galaxies can neither be considered a subset
of typical red galaxies nor a subset of typical blue galaxies, because their
spatial distribution is inconsistent with either one of them. 

It is possible that the dusty red star-forming galaxies may be in the process of
transformation from typical blue field galaxies to typical red cluster galaxies.
But it is equally conceivable that the newly identified population is undergoing
just a particular phase in a more complicated evolutionary picture. 
Miller \& Owen (2002) found star-forming galaxies with large amounts of 
extinction to be concentrated more centrally in clusters than star-forming 
galaxies in general, suggesting an environmental effect. Coia et al. 
(2005) have studied dusty starbursts in several clusters and found great variety 
in abundance even when clusters of similar mass and distance were investigated. 
They found dusty starbursts most often in clusters with complex dynamics that 
were suggestive of sub-cluster mergers. In fact, our target Abell 901/902 is a 
dynamically very complex supercluster environment.

A variety of physical processes are expected to play an important role in the
environmentally triggered transformation from star-forming galaxies to red old
spheroidal systems: e.g., major mergers between galaxies of comparable size 
\citep{Ba92}; impulsive encounters where galaxies move at too high a relative 
speed to merge yet are dynamically heated and sometimes eventually disrupted
(minor interactions and galaxy harassment; Moore et al. 1999); ram-pressure
stripping, where a galaxy's gas content is stripped off as it passes through
the hot, extended gas characteristic of large groups and galaxy clusters
\citep{GG72}; and suffocation, where a galaxy loses its warm gas envelope as
a reservoir for future cooling and gas accretion \citep{LTC80}. Bekki (1999) 
suggested that tidal forces exerted by a cluster onto an infalling group could 
trigger nuclear starbursts with no merging being involved.

Recent results from a variety of authors indicate that the transformation of 
massive star-forming galaxies into spheroidal non-star forming galaxies takes 
place in galaxy groups, rather than in cluster environments \citep{Ko01,Ba02}. 
Therefore, it is necessary not just to study the cores of massive clusters but 
to explore the properties of galaxies out to large radii, towards densities 
characteristic of the field. However, detailed studies of galaxy populations 
in clusters have been very challenging, because without exhaustive spectroscopy 
it is difficult to identify which galaxies are truly embedded in the cluster 
environment.

An alternative route presented here for the first time involves the use of 
high-precision photometric redshifts such as those provided by the medium-band 
survey COMBO-17 \citep{WMR01}. These permit a sufficiently clean selection of 
cluster galaxies with very little field contamination. The detailed medium-band 
SEDs contain a wealth of spectral information which allows simultaneous decoupling
of redshift, age and dust reddening using purely optical photometric data.

One of the COMBO-17 fields contains a complex supercluster environment which
is very suitable to study the evolutionary phenomena near galaxy clusters and
test the role of the physical mechanisms listed above. This is the {\it A901
field} of COMBO-17 which contains the clusters Abell 901a, 901b and 902, as 
well as some associated groups, all at redshift $\sim 0.165$ and within a
projected area of $3.5\times 3.5$~Mpc/$h$. This field has already been subject 
of weak lensing studies in two and three dimensions using the COMBO-17 data 
\citep{Gray02,Tay04}. A first study of the cluster population itself 
investigated the relation of galaxy colours to the dark-matter density as 
inferred from weak lensing maps \citep{Gray04}. The field has since been 
observed with XMM-Newton, {\it Spitzer}, HST/ACS and GALEX.

In this paper, we examine the galaxy population in this cluster complex using 
the optical data from COMBO-17 and taking advantage of accurate photometric 
redshifts which have errors below 0.01 for most objects studied. We exploit the 
17-filter SED data to differentiate between dust-free and obscured galaxies, 
and thus identify dusty galaxies in the red-sequence from our multi-band 
photometry alone. As a result, we are able to identify the true old population
unambiguously, and investigate the properties of the dusty red-sequence 
galaxies. The full COMBO-17 data set provides us with a field comparison.

The paper is structured as follows:
In Sect.~2, we present the COMBO-17 data and their characteristics. Sect.~3
reports on the galaxy sample selected from the COMBO-17 survey catalogue. In
Sect.~4 we compile our analysis of the cluster population characterizing the
several types of galaxies and deriving a type-density relation. Sect.~5 
presents a discussion of our results, and finally we summarize in Sect.~6.

Throughout the paper we use $H_0 = h~\times$ 100~km/(s~Mpc) in combination 
with $(\Omega_{\rm m},\Omega_\Lambda)=(0.3,0.7)$ and use $h=1$ when citing 
luminosities in the text. Both apparent and absolute magnitudes are reported 
in the Vega system and are total object magnitudes. All rest-frame quantities 
are corrected for Galactic foreground reddening, while apparent magnitudes and 
observed-frame colours are used as observed. Rest-frame colour indices are 
always labelled as such, while observed-frame indices have no special labels.

\section{Data}

\subsection{The multi-band SEDs from COMBO-17}

The COMBO-17 survey has measured accurate photometric redshifts for more 
than 30,000 galaxies in three different fields, including the Chandra Deep 
Field South \citep{Wolf04} and the field of Abell 901/902 which is analysed 
in this paper. Photometry in 5 broad (UBVRI) and 12 medium-bands is used to 
classify objects into normal stars, white dwarfs, galaxies and QSOs (more 
generally type-1 AGN) and to estimate redshifts for extragalactic objects.
Quasars are clearly identified when the contribution of AGN light to the 
SED is sufficient to leave recognizable traces in the 17-band SED data.
Comparisons with spectroscopic redshifts and simulations have suggested the 
mean error of COMBO-17 redshifts to be $\sigma_z/(1+z)<0.01$ for galaxies 
with $R<20$, while fainter galaxies have higher redshift errors. We found 
the 90\% completeness limit for galaxies at $z<1$ to be $R \approx 23$, 
fainter than the samples analyzed later in this paper.

The COMBO-17 filters have central wavelengths ranging from observed-frame  
365 to 915~nm and allow us to derive luminosities for cluster galaxies in 
A901 between 310~nm and 740~nm rest-frame wavelength directly from observed 
photometry. We obtain luminosities by placing the best-fitting template SED 
into the observed 17-filter spectrum and integrating the template over the 
efficiency curve of the desired redshifted rest-frame band. However, the 
analysis here relies mostly on observed-band photometry to eliminate impacts
of errors in the derivation of rest-frame quantities. All luminosities refer
to total object photometry, while the colours and SED shapes are determined
from seeing-adaptive aperture photometry probing the same physical section of 
any object in all 17 bands. The faintest objects in the sample contain $\sim 
60$\% of their total light in the apertures, but the few most luminous and 
largest galaxies emit only $\sim 20$\% of the total in the aperture. As a 
result, the SEDs of larger galaxies are more dominated by nuclear light. 

All survey details regarding observations, data reduction, classification,
redshift estimation, rest-frame properties, completeness, accuracy and 
reliability of the object catalogue data are discussed in great detail in 
Wolf et al. (2004). Here, we only repeat a discussion of the galaxy templates
as this is relevant for the analysis of the galaxy population in Abell 901/902.

\subsection{Galaxy templates and SED parameters}

The galaxy template library used in COMBO-17 contains a grid of synthetic 
spectra based on the PEGASE code \cite{FRV97} for population synthesis models. 
The templates span a two-dimensional grid with a range of ages calculated 
by the PEGASE code and a range of extinction levels which we applied as 
external screens to the SEDs delivered by PEGASE. The setup for PEGASE uses 
standard parameters suggested by the codes {\it SSPs} and {\it scenarios} 
with a Kroupa (1993) IMF and no extinction. The star formation history 
follows an exponential decay law with a time constant of $\tau=1$~Gyr. The 
SEDs are calculated by PEGASE for various time steps since the beginning of 
the first star formation. As templates we use 60 spectra which span ages
(more precisely, time since onset of star formation) from 50~Myr to 15~Gyr.
All templates are then extinguished with six different equidistant degrees 
of reddening in the interval of $E_{B-V}=[0.0,0.1,...0.5]$. The reddening
law of our choice is the SMC law from the 3-component model by Pei (1992). 
We decided to use this law because it provided a reasonably good match 
between the rest-frame UV templates and the observed SEDs of galaxies with 
known redshift. Note that the detailed choice of extinction law has no 
effect on the observed SEDs of galaxies with $z<0.6$, and hence certainly
not on the cluster A901.

We tune initial model metallicities to give almost solar metallicity for old
populations typical of local L* spheroidal galaxies. It is worth noting, 
however, that the well-known age/metallicity degeneracy is actually helpful 
in this case: Mismatches between real galaxy and template metallicities can 
be compensated for by modest changes in template age, while yielding still 
accurate estimates of redshift and SED shape.

\subsection{The Abell 901/902 field}

The COMBO-17 observations cover an area of $31\farcm 5 \times 29\farcm 7$ 
at full depth, centered on $(\alpha,\delta)_{2000} = (9^h 56^m 17\fs 3,
-10\degr 01\arcmin 11\arcsec$), which is $(l,b)_{\rm gal} = (248,+34)$. 
The interstellar foreground reddening is estimated at $E_{B-V}=0.06$ 
\citep{SFD98}, which is taken into account for fitting templates to the 
observed photometry and deriving the correct photometric redshift.

At the cluster redshift $z=0.165$, the angular scale is 2.0/$h$~kpc per 
arcsec. Assuming $h=0.7$ the full field covers an area of $5.35 \times 
5.05$~Mpc. This area contains three pronounced cluster cores called Abell 
901a, 901b and 902, as well as various associated groups. A deep WFI 
image of the field has been published in Gray et al. (2002).

The field contains 11009 objects brighter than $R=23$, of which 98\% are 
classified by COMBO-17. Due to its relatively low galactic latitude of $
\sim 34\degr$, the field contains a large proportion of stars. The COMBO-17 
classifier found 2091 stars and 13 white dwarfs at $R<23$. A number of
bright stars reduce the effective field slightly and constrain some 
multi-wavelength follow-up observations. The brightest star in the field 
is an F8 star with $V=9.6$ named PPM 192892=SAO 137231=BD-09 2962. At 
$\lambda>1\mu$ the brightest object is an M8 Mira variable 
\citep[IRAS 09540-0946 = GSC 05479-01188]{KHI97} with $V\approx 15$ 
and $K=5.75$ (measured by 2MASS).

After eliminating stars, QSOs, objects too strange to be classified and 
objects with bad flags, we have 7992 galaxies left with estimated redshifts
and $R<23$ to search for members of the cluster A901/902.

\subsection{Spectroscopy}

Spectra of supercluster galaxies were obtained using the 2dF instrument on 
the AAT in March 2002 and March 2003. Details of these observations are to
be published in Gray et al. (in preparation). A total of 89 galaxies 
were targetted using the 1200B grating in a single fibre configuration 
during the 2002 run, of which 64 were within $0.15<z<0.18$. Three fibre 
configurations using the lower resolution 600V grating were obtained during 
the 2003 run targetting 368 objects, with 47 repeated from the previous run.  
The primary selection function targetted galaxies selected by photometric 
redshift to be within the supercluster redshift and having $R<20$, with 
additional fibres being allocated to secondary targets (QSOs or non-cluster 
galaxies) when available. However, the limitations on close packing of 
fibres meant that the highest-density regions of the clusters were less
well sampled, thereby missing a number of the brightest galaxies. The 
fibers measure $2\arcsec$ in diameter and measure light from very similar 
fractions of any object to the photometric apertures.

Redshifts were obtained in two independent means by line profile fitting
of the the Ca H and K feature in absorption and by cross-correlation with
template spectra using the XCSAO task within IRAF.  The two measures agreed 
well, with $\langle \Delta_z \rangle = 0.00149 \pm 0.00006$. In total, 
spectra were obtained for 407 unique objects including various stars and
QSOs that were targets of different interest. After eliminating several 
galaxies due to data quality issues, we have 249 galaxy spectra 
within the redshift range of the supercluster for statistical analysis.

\begin{figure}
\centering
\includegraphics[clip,angle=270,width=\hsize]{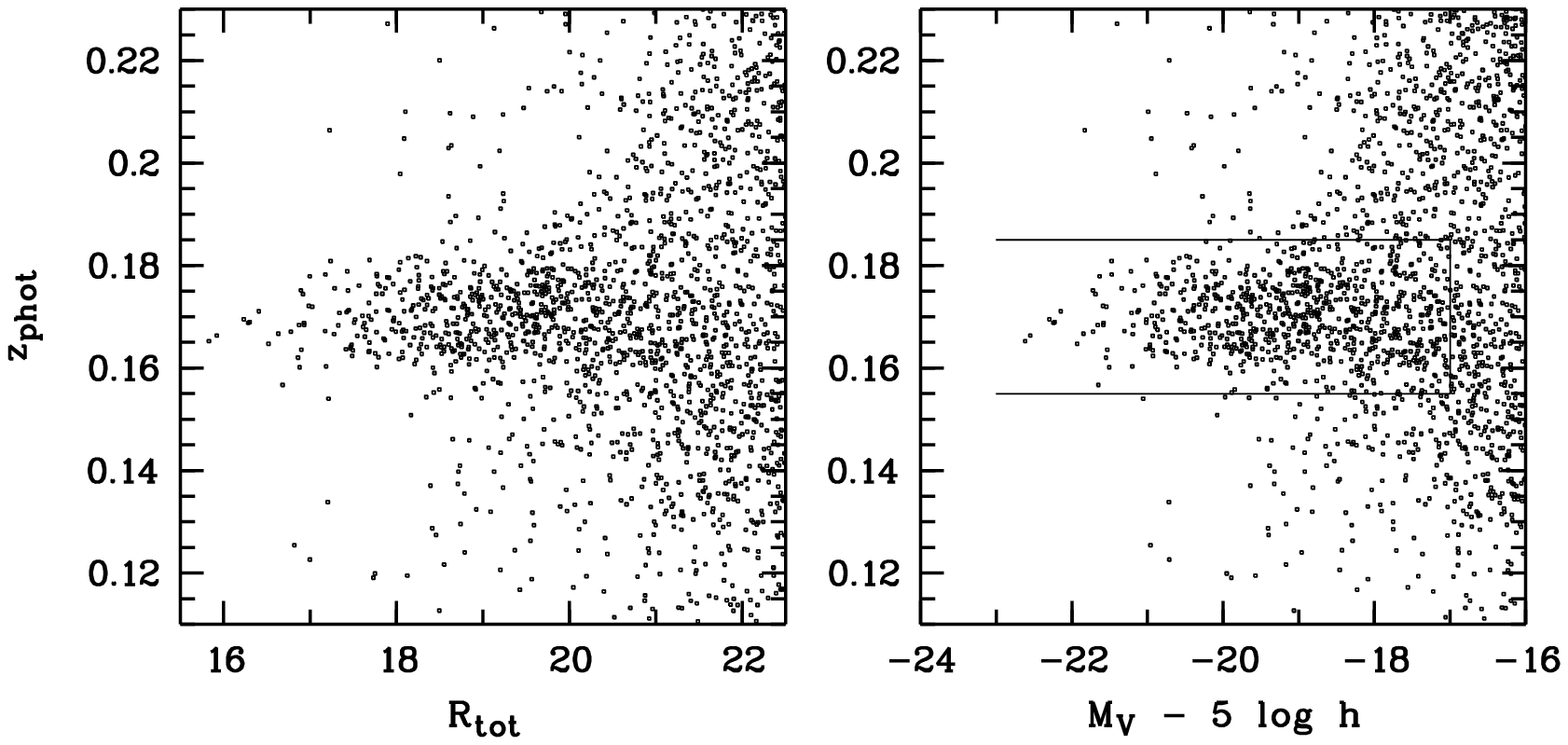}
\includegraphics[clip,angle=270,width=\hsize]{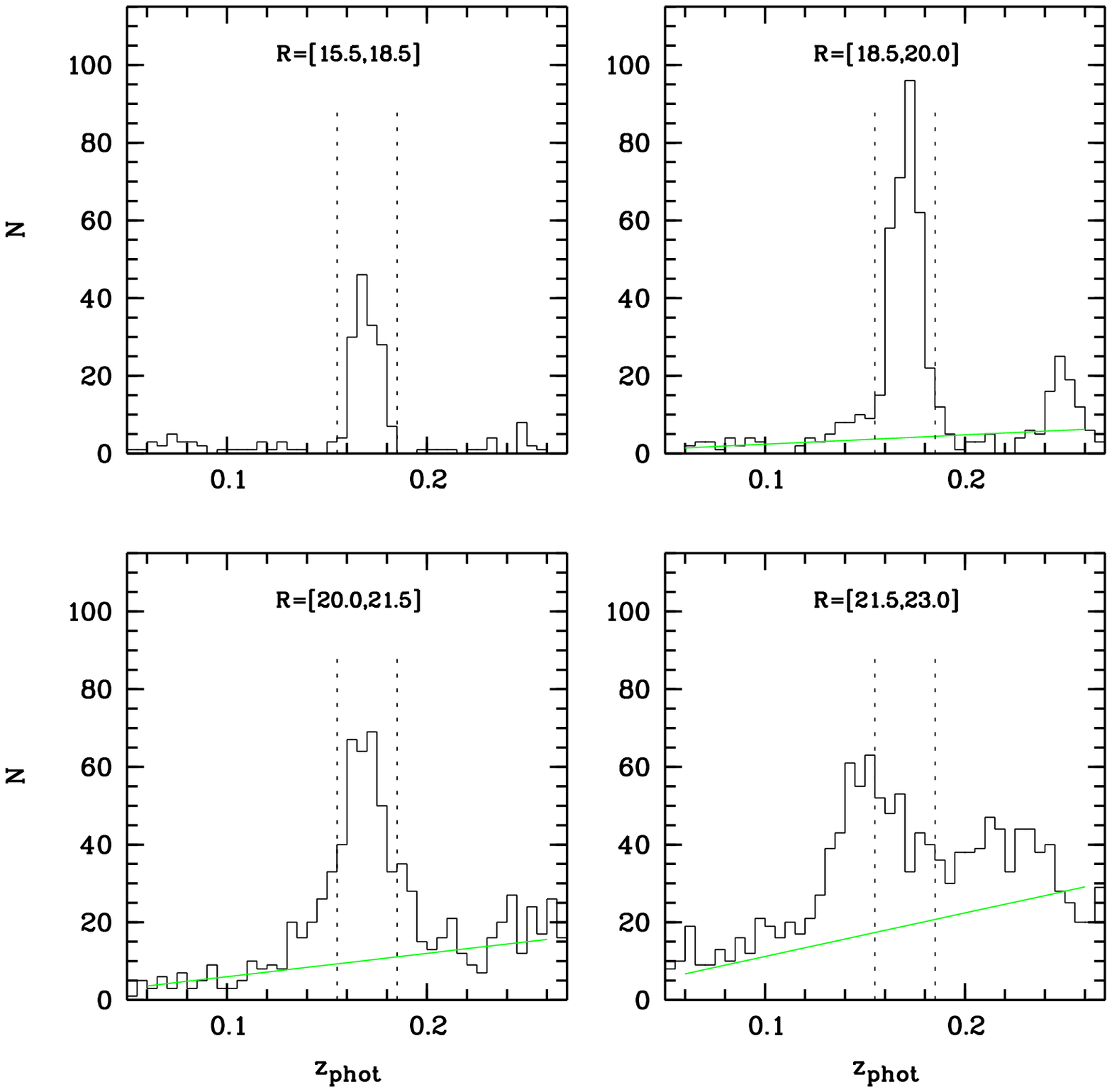}
\caption{Bright cluster galaxies: 
{\it Top row, left:} Photometric redshift vs. apparent magnitude for the 
galaxy sample in the A901 field with the cluster at $z\approx 0.17$.
{\it Top row, right:} Redshift vs. absolute magnitude. The lines show the 
selection for the cluster sample used in the following analysis.
{\it Bottom four panels:} Redshift histograms in apparent magnitude bins.
The dashed lines show the redshift interval for the adopted cluster sample.
At $R>21$ the errors in photometric redshift start to broaden the cluster
distribution and bias it to lower redshift. A rough estimate for the field
contribution is shown in grey.
\label{sampledef}}
\end{figure}

\begin{figure}
\centering
\includegraphics[clip,angle=270,width=\hsize]{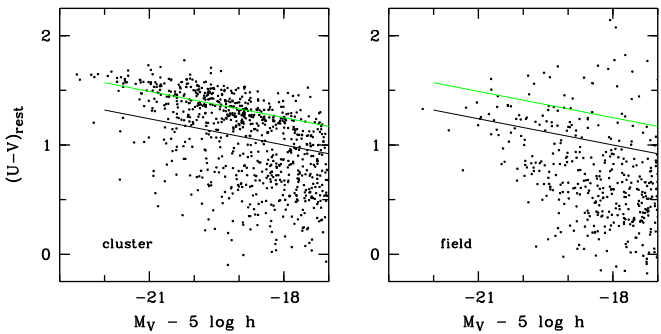}
\caption{Rest-frame colour-magnitude diagrams:
Cluster (left) versus field (right) with cluster red-sequence fit (grey line)
and Butcher-Oemler style red-sequence cut (black line).
\label{CMD1}}
\end{figure}

\section{The galaxy sample}

\subsection{Selecting the cluster population at $M_V<-17$}

Fig.~\ref{sampledef} shows a redshift-magnitude diagram of the galaxy sample
focussing on a redshift interval encompassing the target clusters. The bulk
of the bright cluster population can be clearly seen at $z\approx 0.17$,
reaching up to apparent magnitudes of $R\approx 16$. Fainter than $R=20$ the
photometric redshift errors slowly start to widen the cluster distribution, 
which makes it harder to disentangle cluster members from field contaminants.
In fact, at the bright end ($R<20$) we expect a photometric redshift error of
$\sigma_z <0.01$, while at $R=23$ we expect $\sigma_z \approx 0.03$.

We first decide on a redshift cut for the cluster sample by looking at the 
redshift distribution in a bright sample with small photo-z errors, before
defining a deeper sample on the basis of this redshift cut. For this, we use 
an iterative approach to determine the mean and rms redshift of the cluster 
population. We settled on a redshift cut of $z_{\rm phot}=[0.155,0.185]$ at 
$R<20$. This sample contains 482 galaxies with a redshift distribution of

\begin{equation}
	\langle z_{\rm phot} \rangle = 0.1703 \pm 0.0063   ~,
\end{equation}

suggesting 99\% completeness within the redshift cut if the distribution is 
Gaussian. The scatter translates into a velocity dispersion from photometric 
redshifts of $\sigma_{cz} \approx 1900$~km/s, which is a combination of velocity 
scatter among cluster galaxies and photo-z errors. The velocity scatter in 
the spectroscopic sample is $\sim 1250$~km/s, and requires convolution with 
a photo-z error of $\sigma_z/(1+z)=0.004$ to produce the full scatter. While
this is only the relative redshift scatter within the cluster, the full photo-z 
accuracy is slightly lower: We find the mean redshift in the spectroscopic 
sample to be 0.165, which is offset by 0.005 from the mean photo-z (see also
Fig.~\ref{p3_gals}).

We now define a deeper cluster sample for the following analysis and keep the
redshift range at $z_{\rm phot}=[0.155,0.185]$. The fainter we go, the lower
the completeness of the cluster sample will be as more and more galaxies are
scattered outside the redshift interval due to increasing photo-z errors.
However, widening the interval would clearly include a large amount of field 
contaminants, which we would like to avoid. For this paper, we settle to
applying a luminosity cut at $M_V<-17$, which corresponds to $R\approx 21.5$. 
The low redshift errors in this bright regime help to keep field contamination 
low and cluster completeness high. At $R=21.5$ we expect redshift errors of
$\sigma_z\approx 0.015$, which would make the sample still 68\% complete at 
its luminosity limit. 

Hereby, we select 795 galaxies in a volume of $\sim 1575$~(Mpc/$h$)$^3$. 
Although the clusters occupy only a tiny fraction of this volume, the sample 
consists mostly of cluster galaxies due to their large overdensity. We split 
the sample into red-sequence and blue cloud following the procedure explored 
by Bell et al. (2004) on COMBO-17 galaxies. We apply a cut parallel to the 
colour-magnitude relation of the red-sequence, which evolves with redshift. 
From the COMBO-17 sample Bell et al. (2004) have suggested two alternative 
cuts, one determined internally from COMBO-17 and another one which includes 
measurements from the local Universe. Here, we use the COMBO-17 internal cut 
defined by 
\begin{equation}
	(U-V)_{\rm rest} > (U-V)_{\rm CMR}-0.25 ~,
\end{equation}
where 
\begin{equation}
	(U-V)_{\rm CMR} = 1.48-0.4z-0.08 (M_V-5 \log h + 20) ~. 
\end{equation}
This CMR fit matches the red-sequence of A901/902 at $z=0.165$, 
and splits the cluster sample into 462 red-sequence galaxies and 333 blue 
cloud galaxies (see Fig.~\ref{CMD1}).

\begin{figure}
\centering
\includegraphics[clip,angle=270,width=\hsize]{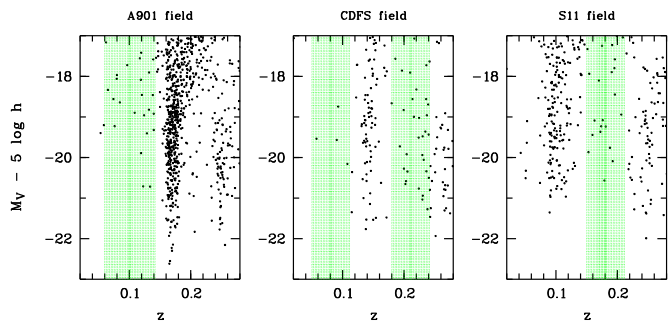}
\caption{Selection of field galaxy sample: Luminosity-redshift diagrams of 
red-sequence galaxies show rich groups and clusters in all three COMBO-17 
fields. We exclude them to select a field sample (shaded area) from the 
low-redshift domain of all fields.
\label{fieldsel}}
\end{figure}

\begin{figure*}
\centering
\includegraphics[clip,angle=270,width=0.75\hsize]{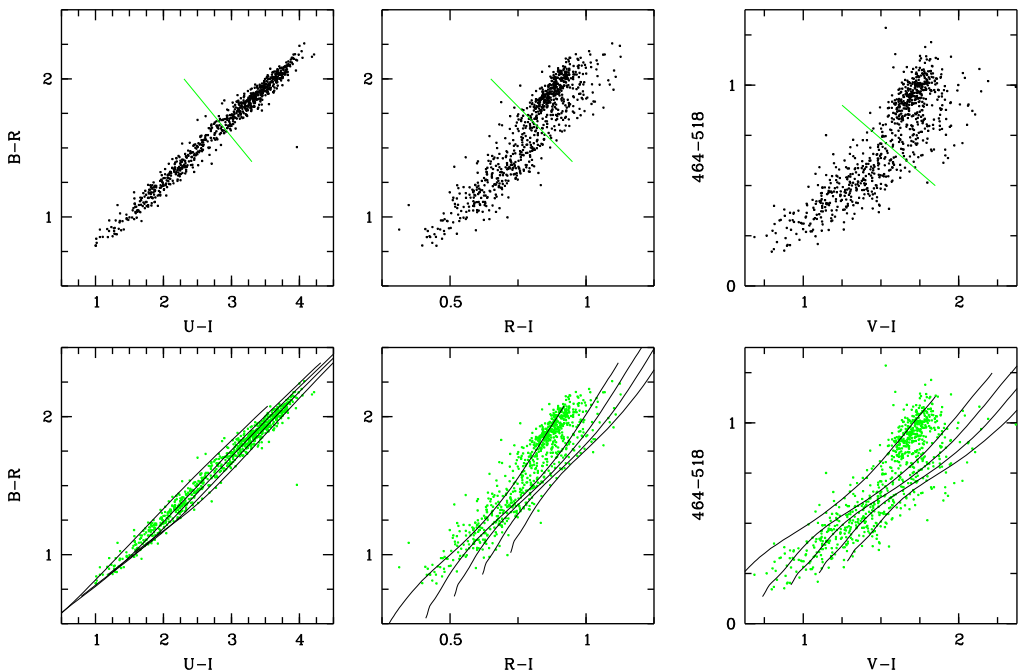}
\caption{Observed-frame colour-colour digrams: 
{\it Top row, left:} Two observed-frame colour-indices, each straddling the 
4000\AA -break, show galaxies as a 1-parameter family. The grey line roughly
indicates the location of the gap in the colour bimodality. 
{\it Top row, center:} When one colour index does not encompass the 4000\AA
-break, the galaxies fan out into a wider range.
{\it Top row, right:} A medium-band colour index probing the 4000\AA -break.
{\it Bottom row:} Template colours (black lines) are plotted over the 
galaxy sample (grey points). The left-most line is an age sequence from 
50~Myr (bottom left) to 15~Gyr (top right) without dust reddening. The 
four other lines show colours for the same age sequence with reddening of
$E_{B-V}=[0.2,0.3,0.4,0.5]$ omitting the line at 0.1 for clarity.
\label{BB_cols}}
\end{figure*}

\subsection{The field sample and field contamination}\label{contamination}

We select a low-redshift field comparison sample for two purposes: we would
like to estimate the field contamination in the redshift slice of the cluster
sample from the space density of galaxies outside the cluster using the same
luminosity cut; and we would like to compare the properties of the cluster 
population to that of the field. 

We would like the field sample to have a similar mean redshift as the cluster 
to suppress possible influence from any evolutionary trends and consider all
three COMBO-17 fields. We investigate the redshift range of $z=[0.05,0.25]$, 
where the photo-z catalogue is more than 90\% complete at $M_V=-17$ for all
galaxy SED types. We use the same luminosity cut and determine the number of 
galaxies and the volume they are drawn from, hence obtaining a space density.

We want to exclude strong local overdensities, such as the Abell clusters at 
$z\sim 0.11$ in the S11 field and at $z\sim 0.17$ in the A901 field as well as 
some rich groups at $z\sim 0.25$. Fig.~\ref{fieldsel} shows overdensities in
redshift space for all three COMBO-17 fields as indicated only by the strongly
clustered red-sequence galaxies. Avoiding obvious overdensities we adopt the 
redshift ranges shaded in grey. The field sample has a mean redshift of 
$\langle z_{\rm field} \rangle =0.172$, which is very close to the redshift 
of A901. It contains 385 galaxies from a volume of $\sim 9750$~(Mpc/$h$)$^3$.

The volume of the cluster sample at $z=[0.155,0.185]$ is more than six times
smaller, and contains an estimated field contamination of 62 galaxies\footnote{
  We note, that e.g. the photo-z selection for cluster galaxies at 
  $z=0.41$ based on pure broad-band photometry in Kodama et al. (2001) used a 
  5.33 times wider redshift bin, which led to an 33 times higher estimated field 
  contamination: the physical size of their studied area was $\sim 3\times$
  larger as well, and the luminosity cut was marginally fainter.}.
This implies that statistically one in 13 out of the 795 galaxies in the cluster 
sample would really be in the physically unrelated field, somewhere along the 
line-of sight within $\pm 50/h$~Mpc from the cluster center. Since red-sequence 
galaxies are intrinsically more clustered than blue galaxies, we split the 
contamination assessment by galaxy colour by applying the evolving red-sequence 
cut to the field sample. Our best estimate is that 13 out of 468 red-sequence 
galaxies (3\%), and 49 out of 327 blue-cloud galaxies (15\%) are field 
contaminants. No attempt is made in this paper to subtract the field 
contamination from any numbers reported for the cluster.

\subsection{Galaxies as a 2-parameter family}

To first order, galaxy SEDs are often considered to form a 1-parameter family,
running from old populations in red spheroidal galaxies to young starbursts in 
blue galaxies \citep{Con95}. Also, plenty of photometric redshift codes use a 
1-parameter family of templates. In COMBO-17, such an approach was taken until 
2002, by when it became clear that this was inappropriate at least for the 
galaxies in A901. Using the Kinney et al. (1996) template spectra, $\sim 1/4$ 
of the bright galaxies in the cluster had redshift errors up to $\la 0.1$ and 
would have been left out from the study presented here.

\begin{figure*}
\centering
\includegraphics[clip,angle=270,width=0.75\hsize]{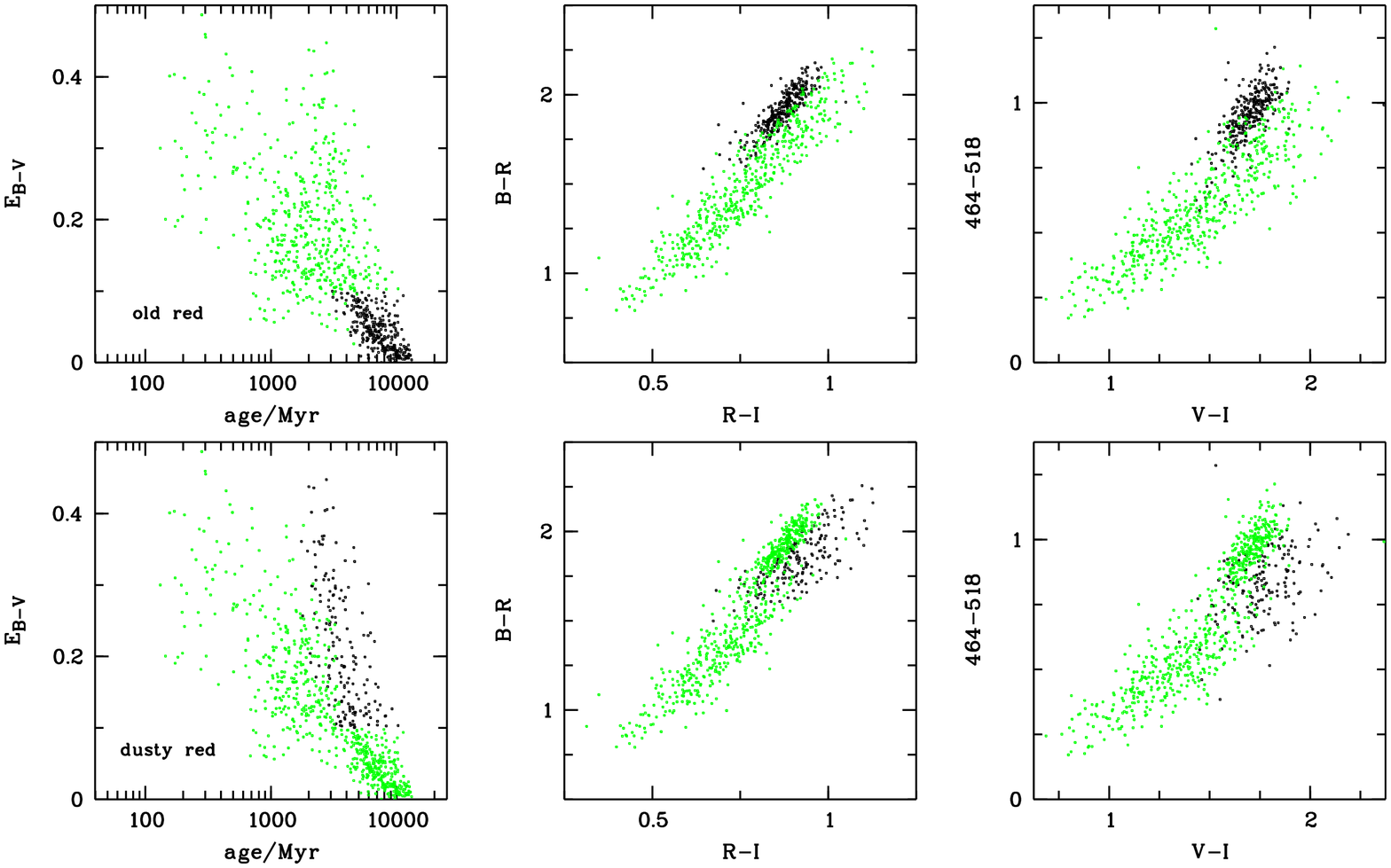}
\includegraphics[clip,angle=270,width=0.75\hsize]{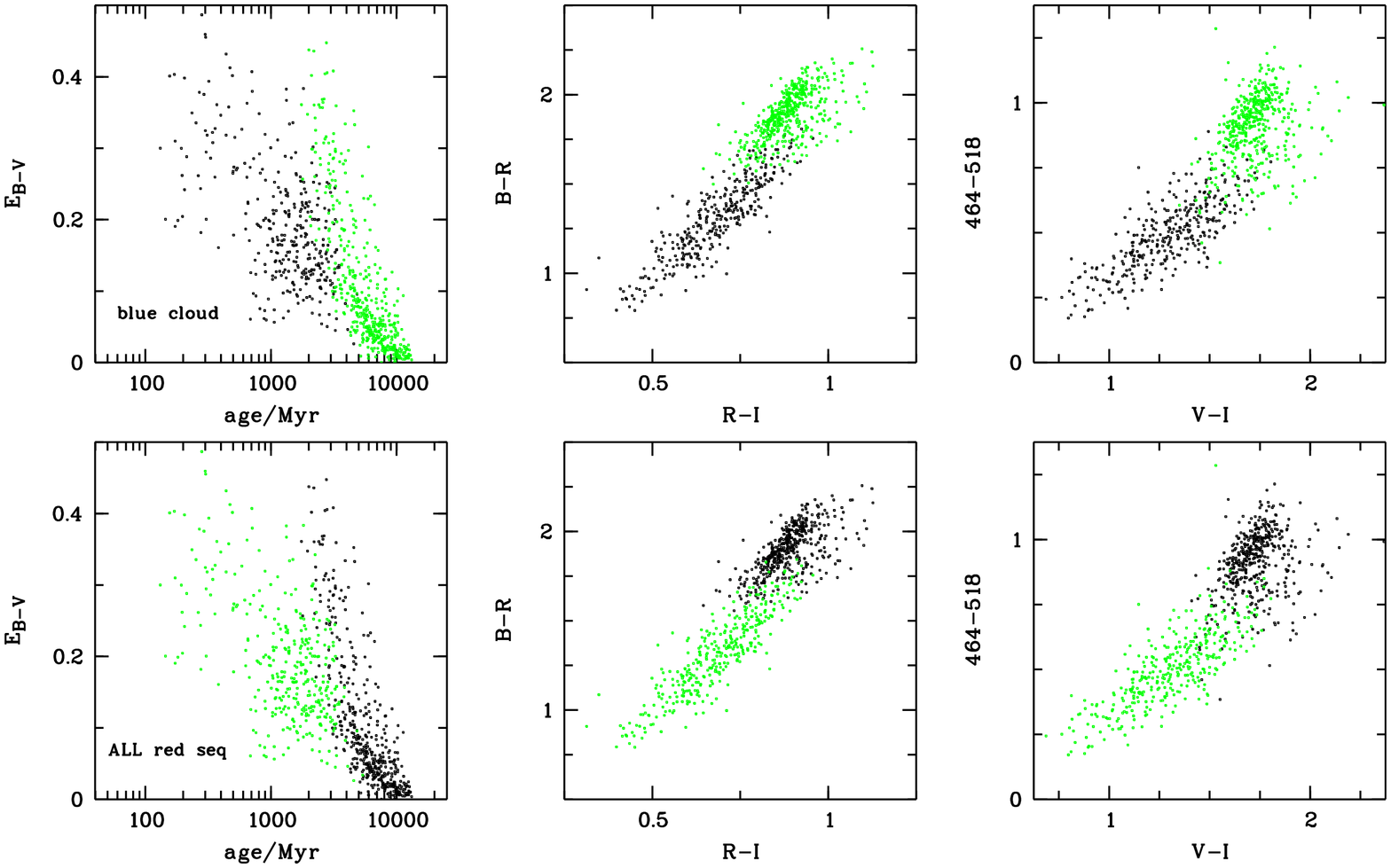}
\caption{Three galaxy types: 
{\it Left:} Stellar age vs. $E_{B-V}$ from estimated template parameters. 
{\it Center/Right:} Broad-band/medium-band colour. 
{\it Top row:} Dust-free old clump in the red-sequence.
{\it Middle row:} Dusty population in the red-sequence.
{\it Bottom row:} The blue cloud.
\label{clump_sel}}
\end{figure*}

However, using the 2-D template set outlined above the redshifts of virtually 
all cluster galaxies were correctly estimated. The 2-parameter aspect of the
galaxy family is of central importance, to accommodate spectra with different
curvature while keeping the same overall colour. This issue becomes apparent 
when looking at colour-colour diagrams of the cluster in Fig.~\ref{BB_cols}.
Here, the panels on the left-hand side show a narrow distribution of galaxy
colours in observed-frame $B-R$ over $U-I$. The approximate position of the 
gap between red-sequence and blue cloud galaxies is given by a line cutting 
across the distribution, although the gap is washed out by the CMR slope in
combination with a wide luminosity interval collapsed in this plot. The center
panels show galaxies more spread out in the colour plane $B-R$ over $R-I$, 
and suggest the existence of a distinctive elongated clump at the red end,
almost detached from the main, larger and less concentrated, distribution.

The observed-frame $B-R$ encloses the 4000\AA -break at $z\sim 0.16$ as does
the colour index $U-I$, in contrast to the colour index $R-I$ which probes
the red end of the observed stellar spectrum. The spread in $R-I$ colour at
fixed $B-R$ colour means that the 4000\AA -break is not a safe predictor for
the colour at longer (or even shorter) wavelengths. Since the 4000\AA -break
is mostly driven by the mean stellar age of a galaxy, the spread in colours
requires another important factor.

A comparison with the age\,$\times$\,dust template grid shows dust reddening 
to be a possible second independent factor in galaxy colours. The pronounced 
red clump coincides with templates of old age and no dust reddening, and is
hereafter referred to as {\it dust-free old clump}. It may represent typical 
spheroidal galaxies in the cluster, while galaxies with redder $R-I$ colours 
at fixed $B-R$ value correspond to dust-reddened objects with younger mean 
stellar age. This interpretation of galaxies as a 2-D age\,$\times$\,dust 
family can further be tested with the COMBO-17 medium-band colours. In the 
right-hand panels of Fig.~\ref{BB_cols} we see again the clear 2-D parameter
family. Colour indices from neighboring medium-band filters probe small-scale
variations in the SED. Near the 4000\AA -break these differ between galaxies
that are red because of age and those that are red because of dust. We note,
that medium-band indices where both filters are redwards of the 4000\AA -break 
do not differentiate age and dust, and galaxies appear again as 1-D families.

\begin{table}
\caption{Mean properties of the three galaxy SED class samples. 
\label{class3} }
\begin{tabular}{lrrr}
\hline \noalign{\smallskip} \hline \noalign{\smallskip} 
property	&  dust-free old	& dusty red-seq		& blue cloud  \\ 
\noalign{\smallskip} \hline \noalign{\smallskip}
$N_{\rm galaxy}$		&  294	&  168	&  333  \\
$N_{\rm field\,\, contamination}$	&   6	&   7	&   49  \\  
$N_{\rm spectra}$		&  144	&   69	&   36  \\ 
\noalign{\smallskip} \hline \noalign{\smallskip}
$z_{\rm spec}$		&0.1646	&0.1646	&0.1658 \\  
$\sigma_{cz}/(1+z)$/(km/s)	& 939	& 1181	& 926 \\   
$z_{\rm spec, N}$		&0.1625	&0.1615	& N/A \\  
$z_{\rm spec, S}$		&0.1679	&0.1686	& N/A \\  
$\sigma_{cz,\rm N}/(1+z)$/(km/s)	& 589	& 597	& N/A  \\ 
$\sigma_{cz,\rm S}/(1+z)$/(km/s)	& 522	& 546	& N/A  \\ 
$\log (\Sigma_{10} ({\rm Mpc}/h)^2)$	& 2.188 & 1.991	& 1.999 \\    
EW$_e$ (O{\sc ii})/\AA	& 	N/A	& $4.2\pm 0.4$	& $17.5\pm 1.5$ \\
EW$_a$ (H$\delta$)/\AA	& $2.3\pm 0.5$	& $2.6\pm 0.5$	& $ 4.5\pm 1.0$ \\
\noalign{\smallskip} \hline \noalign{\smallskip}
age/Gyr			&  6.2	&  3.5	&  1.2  \\   
$E_{B-V}$		& 0.044	& 0.212	& 0.193 \\   
$(U-V)_{\rm rest}$	& 1.372	& 1.293	& 0.670 \\  
$M_{V,\rm rest}$	&-19.31	&-19.18	&-18.47 \\
\noalign{\smallskip} \hline \noalign{\smallskip}
$B-R	$	& 1.918	& 1.847	& 1.303 \\
$V-I	$	& 1.701	& 1.780	& 1.290 \\
$R-I	$	& 0.870	& 0.920	& 0.680 \\
$U-420$		& 0.033	&-0.079	&-0.377 \\
$420-464$	& 0.537	& 0.602	& 0.560 \\
$464-518$	& 0.954	& 0.827	& 0.490 \\
$604-646$	& 0.356	& 0.339	& 0.238 \\
$753-815$	& 0.261	& 0.274	& 0.224 \\
\noalign{\smallskip} \hline
\end{tabular}
\end{table}

\subsection{What is in the red sequence? Provisional galaxy SED classification}

The colour-colour diagrams already suggested that we have two contributions 
to the red-sequence of Abell 901/902: a pronounced dust-free old clump; and
a more widely scattered collection of galaxies with higher dust reddening
and lower mean stellar age. We thus try go beyond a simple classification of
galaxies into a red-sequence and a blue cloud from their colour bimodality.
We want to split the red-sequence itself into two different types in order to
investigate their possibly different properties in terms of distribution in
space, luminosity, their environment and their spectral properties.
We try to separate the pronounced dust-free old clump from dusty red-sequence
galaxies by cutting in template parameter space rather than colour space. We 
find ad-hoc that a simple reddening cut at $E_{B-V}=0.1$ delivers a reasonably 
clean separation between the two features (see Fig.~\ref{clump_sel}).

For the purpose of this paper, we will now provisionally classify the galaxy 
population into three categories: 
\begin{enumerate}
\item As usual, on the blue side of the red-sequence cut, we have the {\it blue 
  cloud} of star-forming galaxies.
\item The {\it dust-free old clump} are red-sequence galaxies with $E_{B-V}<0.1$.
  We consider them an improved selection of old stellar populations with much
  less contamination from dust-reddened galaxies than a regular red-sequence cut
  (at least in Abell 901/902).
\item The {\it dusty red-sequence} are red-sequence galaxies with $E_{B-V}>0.1$.
  We consider them dusty contaminants to the red-sequence or an older-age tail
  of the blue cloud reaching across the red-sequence, and we are surprised by 
  their large number. We investigate their nature and relation to the other
  galaxy types further in this paper.
\end{enumerate}

This classification is an improvement over the simple bimodality split of the 
galaxy population and motivated by both an interest in analyzing dust-reddened
contaminants of the red-sequence, and a desire to have a cleaner sample of old,
passively evolving stellar populations. However, it is no ideal classification 
from a physical point of view. Eventually, we would rather want to understand 
how the mean stellar age or the total star formation rate depend on galaxy 
environment. But the total star formation rate can not be measured from optical 
data alone as it does not provide any clues to how much star formation is hidden 
by dust. We are looking forward to ongoing observations with {\it Spitzer} which 
will provide a full account of star formation in this cluster. In this paper, we 
will investigate the properties of the three galaxy classes as outlined above.

\begin{figure}
\centering
\includegraphics[clip,angle=270,width=\hsize]{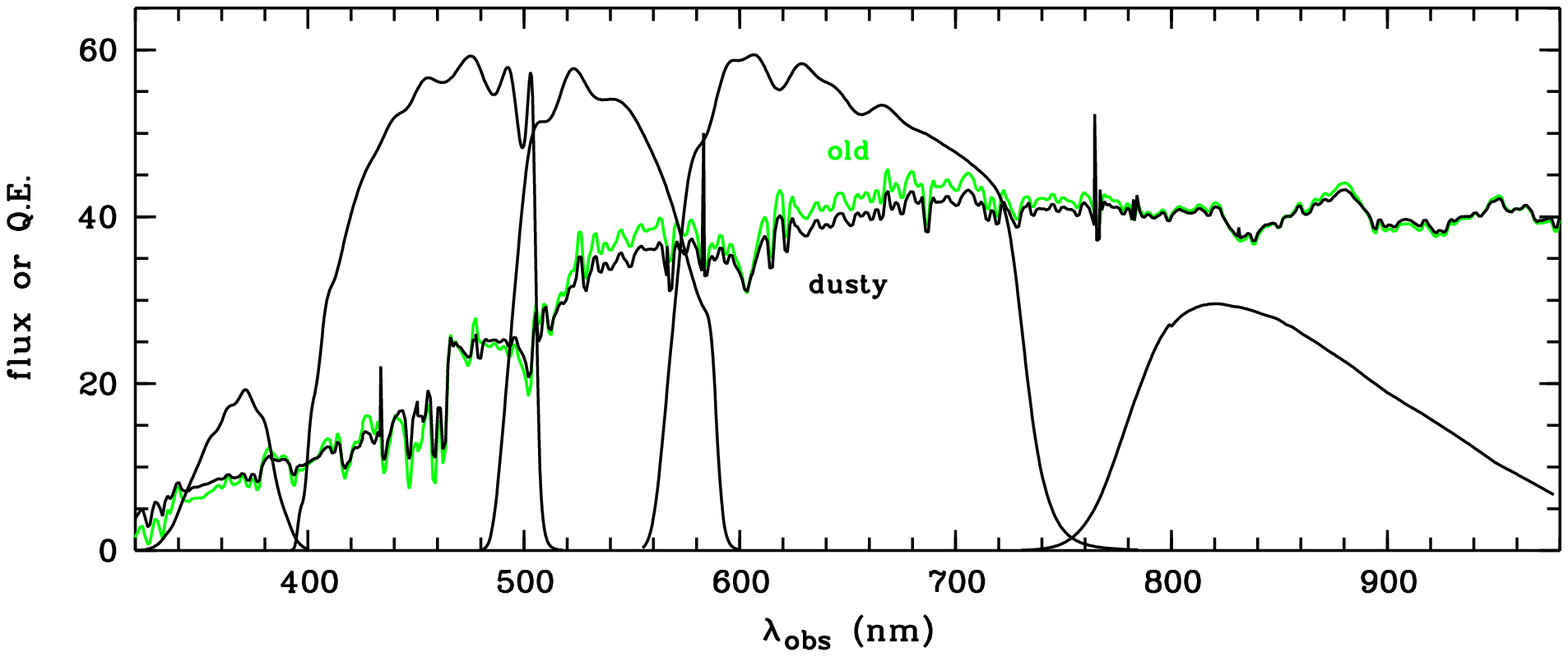}
\includegraphics[clip,angle=270,width=\hsize]{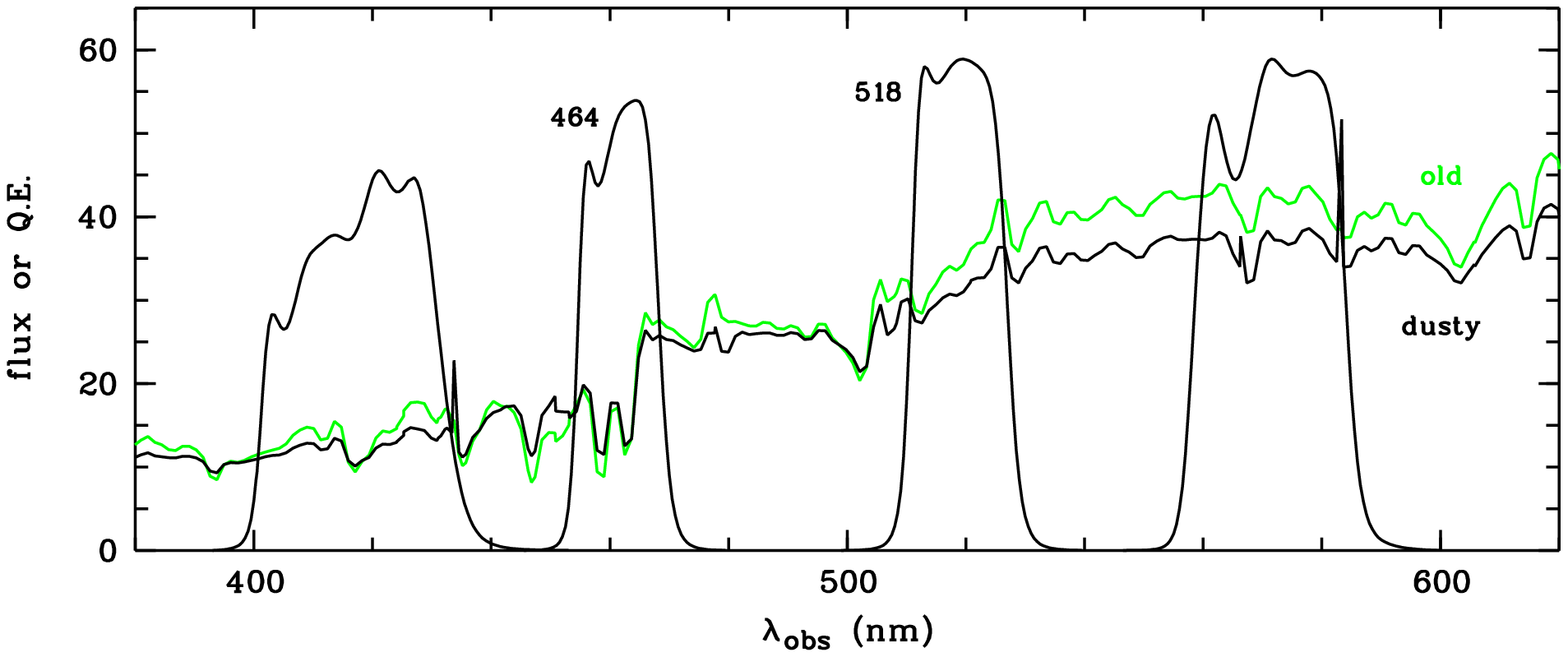}
\includegraphics[clip,angle=270,width=\hsize]{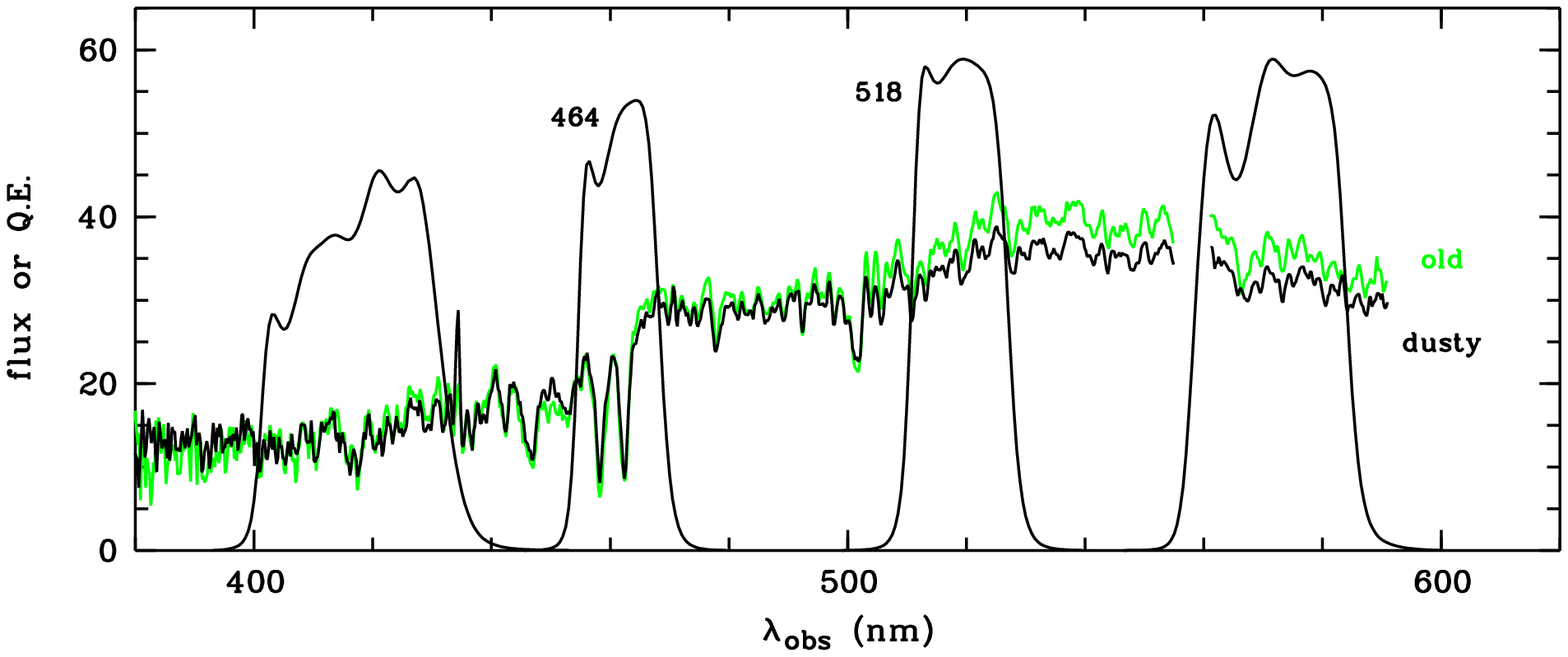}
\caption{Colours and spectra:
{\it Top:} The mean template of the dusty red-sequence galaxies is redder at the
red end and bluer at the blue end than that of the dust-free old clump galaxies.
{\it Middle:} The dust-free old clump reflects older age in a stronger 4000\AA 
-break and Ca H\&K-lines than the dusty red galaxies. The spectral differences in 
these templates match up with the mean colour differences of the two SED classes. 
{\it Bottom:} Similar trends can be found in the mean observed spectra (we 
note that the flux calibration is incorrect towards the red).
\label{templat}}
\end{figure}

\begin{figure}
\centering
\includegraphics[clip,angle=270,width=\hsize]{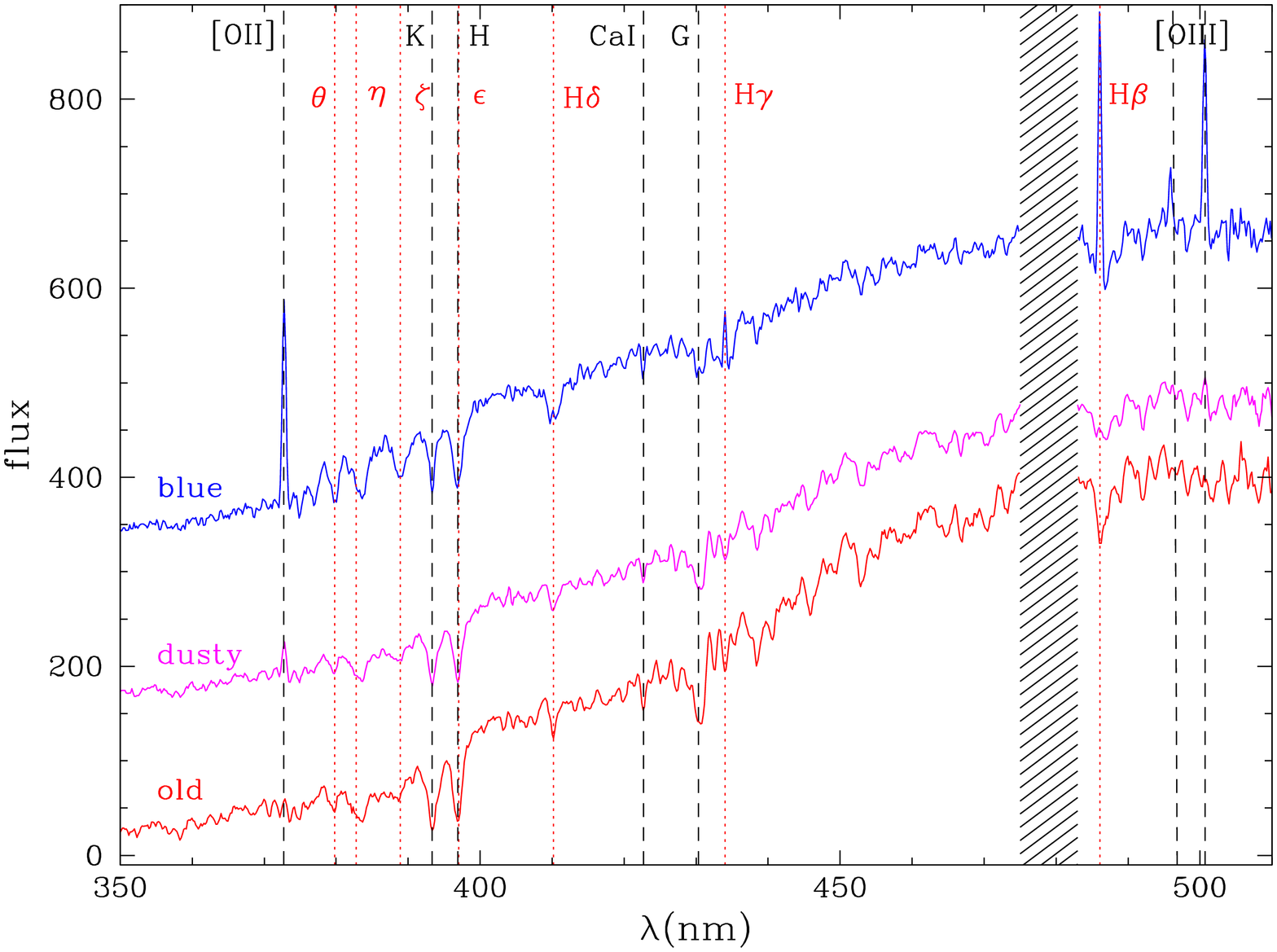}
\includegraphics[clip,angle=270,width=\hsize]{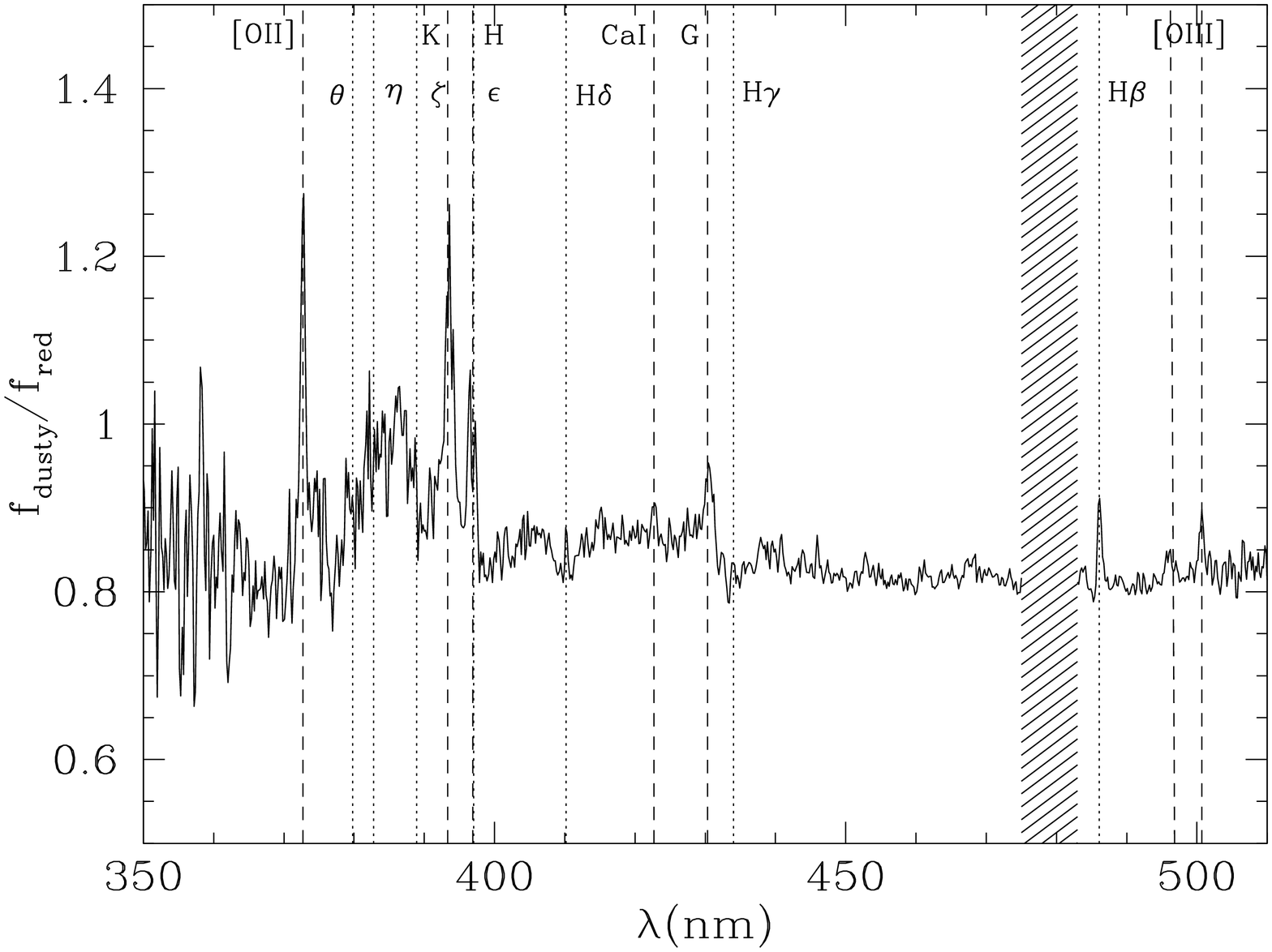}
\caption{
{\it Top:} Mean 2dF spectra for three galaxy types (restframe). Vertical offsets
are used for clarity. The hashed area is affected by night-sky emission lines.
{\it Bottom:} The mean dusty red spectrum divided by the mean old red spectrum.
\label{mean}}
\end{figure}

\section{Analysis of galaxy types}

\subsection{Mean SEDs for three classes}

We would now like to investigate the average spectral properties among the
three galaxy SED classes just defined. We will ignore the age-metallicity 
degeneracy here, since a detailed discussion of ages and star formation 
histories is beyond the scope of this paper, and the degeneracy will not 
matter much for the overall shape of continuum spectra and colour indices. 

From the template parameter values we find mean dust extinction levels and
galaxy ages given our choice of star formation histories and metallicities.
The dust-free old clump is distinguished from dusty red-sequence contaminants 
by a high mean age of 6.2 Gyr and a low extinction of $\langle E_{B-V} \rangle
\approx 0.04$ as opposed to 3.5 Gyr and $\langle E_{B-V} \rangle \approx 0.2$. 
The dusty red-sequence contaminants appear with similar dust extinction levels
as the blue cloud galaxies do, but with a higher galaxy age that could reflect
a larger proportion of old stars. However, there could be potentially stronger 
extinction affecting specifically the youngest stellar population.

In Fig.~\ref{templat} we show the two mean templates for the dust-free and
the dusty red-sequence galaxies along with the full quantum efficiency curves
of several COMBO-17 filters. In the top panel, we can see how both templates
have the same $U-I$ colour, but differ in between those filters. The dusty red
template is bluer at the blue end of the spectrum due to an increase of young
stars. At the long-wavelength end it is redder than the dust-free old clump,
presumably due to dust reddening. Comparing only the slope of the continuum
between neighboring filters reveals only little difference between them. 
However, in the colour-colour diagrams of Fig.~\ref{clump_sel} the two types of 
red galaxies are more clearly differentiated because their principal colour 
difference is orthogonal to the extent of the dust-free old clump.

The middle panel shows template details around the 4000\AA -break and a few
medium-band filters (here filter names are made from their central wavelength). 
Here, the templates are normalised to the flux in the 464-filter, which contains 
the Ca H\&K lines. The different template flux levels within the filters match 
up in fine detail to the different mean colours in the sample: When compared to 
the dusty red galaxies, the dust-free old clump is redder in $U-420$, bluer in 
$420-464$, again redder in $464-518$ and similar in $518-571$. The detailed 
correspondence between template details and sample colours vindicates the use 
of the template grid with its age\,$\times$\,dust structure.

The similarity in broad-band colours between old red and dusty red galaxies
in such a low-redshift cluster echoes a similar ambiguity in interpreting
the colours of Extremely Red Objects (EROs) found at higher redshift. These
are believed to comprise dust-free, old galaxies as well as dust-reddened 
star-forming galaxies, mostly at $z\sim 1\ldots 2$, but also include edge-on
disk galaxies at lower-redshift \citep{YT03}.

\begin{figure*}
\centering
\includegraphics[clip,angle=270,width=0.8\hsize]{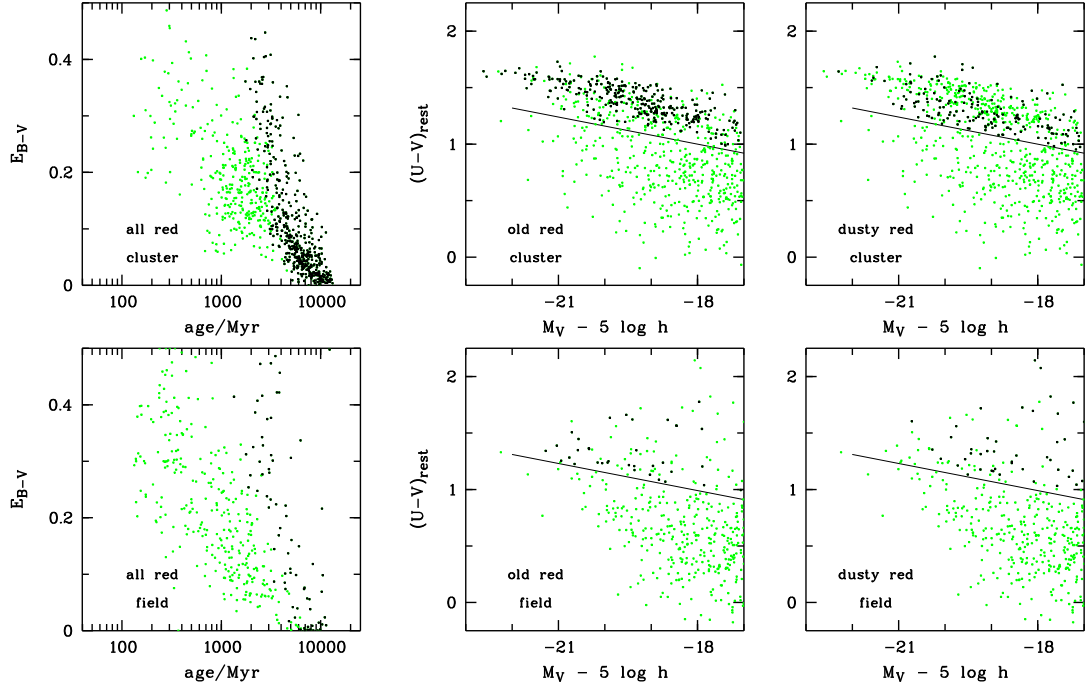}
\caption{Cluster (top) versus field (bottom): 
{\it Left:} The age\,$\times$\,dust-parameter plot shows great enhancement of 
red-sequence galaxies in the cluster, as well as a shortage of young ($<1$~Gyr)
populations. The cluster-specific enhancement applies to both, dust-free old
galaxies ({\it center}) and dusty red-sequence contaminants ({\it right}). The 
number of objects in the cluster is roughly twice that in the field (795:385).
\label{clusterfield}}
\end{figure*}

\subsection{Mean spectra for three classes}

However, in order to deduce the true nature of the dusty red-sequence galaxies
we should investigate their observed spectra rather than the templates. We have
averaged the spectra for the three types using equal weights for every galaxy.
Most spectra fall into the dust-free old category because these galaxies are
most abundant at high luminosity and dominate any bright flux-limited sample.
With 69 objects the dusty red-sequence category contains enough data to allow 
a detailed comparison with galaxies from the dust-free old clump. 

We show the mean spectra in Fig.~\ref{mean} with relative flux offsets for
clarity. From old red over dusty red to blue young galaxies, there is a clear 
trend of increasing Oxygen and Hydrogen emission line fluxes. Going from old
red to dusty red galaxies, emission fill-in can be seen within the H$\beta$ 
and H$\delta$ absorption lines. Going again from old via dusty to blue, the 
Ca absorption lines and the G-band decrease in depth. These trends may appear
quite weak when going from old red to dusty red galaxies but can be seen more
clearly in the ratio spectrum of Fig.~\ref{mean}, where dusty red galaxies are 
divided by old red galaxies, and even weak O{\sc iii} emission lines appear.

We compare the two types of red galaxies more carefully in Fig.~\ref{templat}.
Redwards of the Ca lines, we find again the same trends first identified from 
the medium-band colour index $464-518$ and from the templates. Bluewards of Ca 
H\&K the spectra are relatively noisy and we see less pronounced differences. 
The mean spectrum of dust-free old galaxies reflects an old stellar population 
which is passively evolving and shows no signs of ongoing star formation. 
We see some H$\delta$-absorption which hints at recent star formation. The 
equivalent width of H$\delta$ is 2.3~\AA \ (rest-frame), while normal cluster 
ellipticals are not supposed to show any clear H$\delta$-lines. Measuring lines 
in the mean spectrum should be robust in terms of measuring the average 
equivalent width, but it tells us nothing about the population mix producing this 
average. The mean spectrum may contain a fraction of k+a galaxies (post-starburst 
galaxies also known as E+A galaxies) with pronounced H$\delta$-lines.

In the dusty red-sequence galaxies we see weak O{\sc ii} emission as well as
H$\delta$-absorption{, and conclude that they must be forming stars at some 
rate currently. Given the template solution fitting the observed colour spectrum,
we find that they are moved to the red sequence via a combination of dust reddening 
typical of blue cloud galaxies, together with older ages typical of red-sequence 
galaxies.
We find equivalent widths in emission of $\langle$EW$_e$(O{\sc ii})$\rangle = 
4.2$~\AA \ and in absorption of $\langle$EW$_a$(H$\delta$)$\rangle =2.6$~\AA . 
In contrast, the average blue cloud galaxy has not only stronger Oxygen emission
($\langle$EW$_e$(O{\sc ii})$\rangle =17.5$~\AA ) but also stronger H$\delta$ 
absorption ($\langle$EW$_a$(H$\delta$)$\rangle =4.5$~\AA ).

The equivalent width of H$\delta$ holds clues about the recent variations in
the star formation rate. Values of EW$_a$ (H$\delta$)$ >5$~\AA \ imply that a
recent starburst was followed by quenching of the star formation, such as in
k+a galaxies. If the star formation had continued, related emission would have 
filled in the line reducing the EW (see Mercurio et al. 2004 and references
therein). Dressler et al. (1999) introduced the notion of {\it e(a) galaxies}, 
which display not only strong H$\delta$ absorption as in k+a galaxies, but 
also O{\sc ii} emission suggesting ongoing star formation. These objects are 
explained by Poggianti \& Wu (2000) as dust-enshrouded star bursts where 
age-dependent dust absorption explaines the apparent contradiction from the 
lines: Strong H$\delta$ troughs arise from an A star population that has left 
their dusty birth place, while the very young O stars are still hidden such that 
they can not fill in the absorption with emission from their ionized environment.

At this stage, we will not offer a detailed explanation of the spectra of our 
class of dusty, star-forming red-sequence galaxies, nor speculate about deeply 
dust-enshrouded star formation. A more detailed analysis of individual spectra 
is the subject of another paper (Gray et al., in preparation), and {\it Spitzer}
observations of this cluster field have just been obtained. 

\begin{figure*}
\centering
\includegraphics[clip,angle=270,width=0.8\hsize]{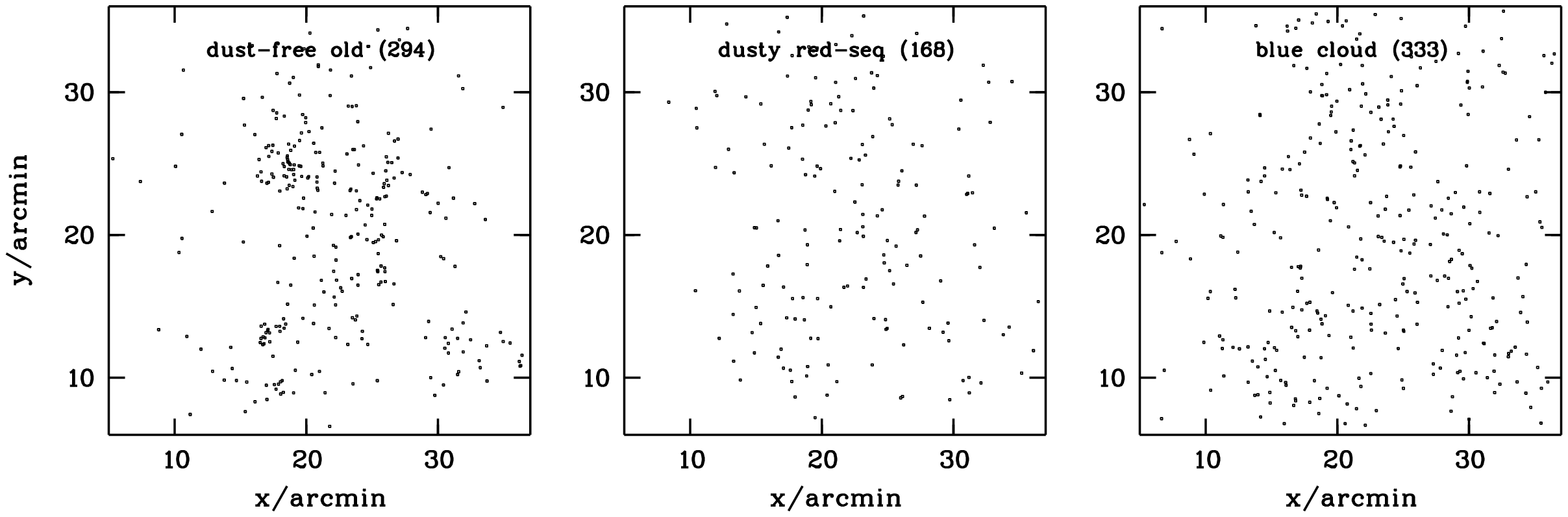}
\includegraphics[clip,angle=270,width=0.8\hsize]{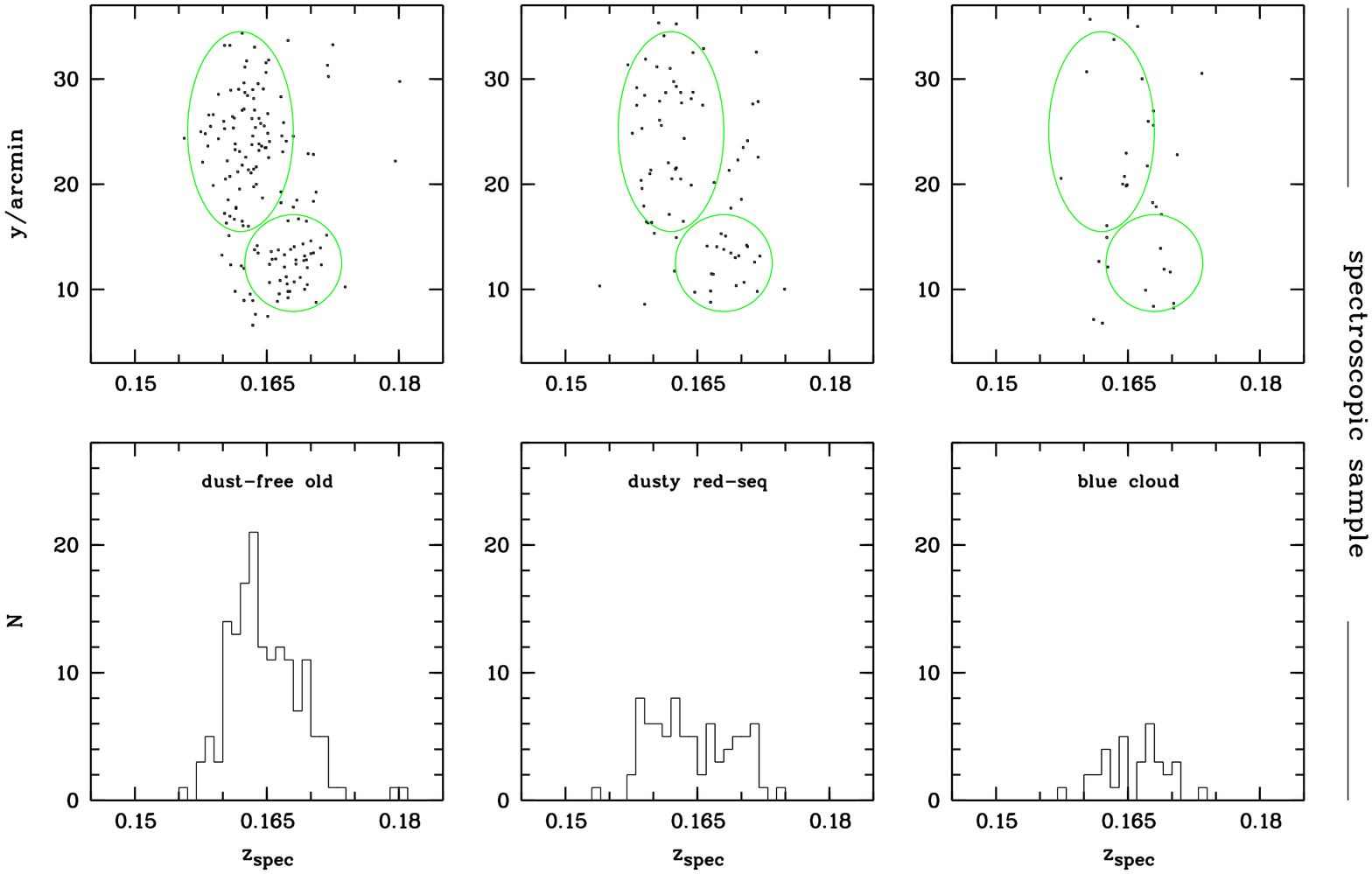}
\caption{Distribution of galaxy types in $(x,y,z)$: 
{\it Top row:} Sky maps. The dust-free old galaxies are concentrated towards
the cluster cores, while the other two types are more scattered over the field.
{\it Middle row:} The cluster cores form two main velocity components, a North
group and a South group separated by $\Delta (cz)/(1+z) \approx 1400$~km/sec in 
rest-frame velocity. The dusty red galaxies have a similar velocity distribution 
as the old red objects. Blue galaxies tend to avoid the locus of red galaxies.
{\it Bottom row:} Redshift distributions reflecting mostly internal dynamics.
\label{xymaps}}
\end{figure*}

\subsection{Cluster vs. field -- contamination?}

We now compare the galaxy properties between the cluster sample and the field
sample in order to assess to what extent field contamination could bias the
cluster results. Especially, we would like to see whether the relatively high  
fraction of intermediate-age dust-reddened galaxies within the red-sequence of 
the cluster is a result of field contamination to the cluster sample.

In Fig.~\ref{clusterfield} we compare age\,$\times$\,dust-parameters and 
colour-magnitude diagrams of the cluster and field samples. We find that 
the cluster contains an enhancement of both dust-free old populations and 
intermediate-age dust-reddened populations (at $\sim 3$~Gyr on our mean 
stellar age scale). On the other hand, the cluster shows a shortage of blue
cloud galaxies, which is particularly apparent among young populations with 
ages of $<1$~Gyr. In the cluster, more than a third of all galaxies belong to 
the dust-free old clump, while in the field only 9\% do. More than 20\% of the 
cluster sample are in the red-sequence despite not belonging to the dust-free 
old clump, while only 11\% of field galaxies do. The dusty red-sequence 
galaxies reach higher luminosities in the cluster than in the field.

In Sect.~\ref{contamination} we reported numbers on field contamination to the
cluster sample as expected from the galaxy densities measured at non-cluster 
redshifts. The result was a considerable contamination to the blue cloud, but 
in the cluster red-sequence only 3\% were expected to be field galaxies. This
contamination rate is much below the observed contribution from dust-reddened
galaxies to the cluster red-sequence. These intermediate-age dust-reddened 
galaxies are therefore a genuine cluster-related phenomenon, which can not be
explained by effects relating to selection or contamination.

While the increased red-sequence fraction and the shortage of young vigorously 
star-forming galaxies in clusters has been established long ago, the increase
in dust-reddened intermediate-age populations is a new trend identified here 
from optical data alone.

\subsection{Spatial clustering and velocity distribution}

The spatial distribution of the three galaxy types in the cluster is shown in
Fig.~\ref{xymaps}. The top row shows sky maps with positions straight from the
camera images. It is immediately clear that galaxies from the dust free old 
clump are most highly clustered, and reach the highest projected densities at 
the cores of the various subclusters. In contrast, dusty red galaxies and blue 
cloud galaxies show no high concentrations within this supercluster environment. 
However, we can decuce from the contamination estimate, that the total density 
of blue galaxies is still six times higher than average field galaxy density.
The maps may suggest to the eye that both dusty red and blue galaxies trace 
roughly the shape of overdensities outlined by the dust-free old population, 
just without their strong clustering. 

Going beyond 2-D projections we try to investigate the 3-D distribution in 
$(x,y,z)$-coordinates, although it is clear, that the $z$-coordinate reflects 
a combination of spatial position with peculiar velocity. The middle row in 
Fig.~\ref{xymaps} shows projected maps in the $(y,z)$-plane (North is up). The
strongly clustered old galaxies reveal two concentrations, a Northern group 
(ellipse) and a Southern group (circle), with a rest-frame velocity difference 
of $\Delta (cz)/(1+z) \approx 1400$~km/sec centered on $cz = 49500$~km/sec. 
There is also indication for 
filamentary structure. The $(y,z)$-maps can not easily show clustering as such, 
because galaxies in deep potential wells show wide velocity dispersions. We note, 
that projections in the $(x,z)$-plane happen to not reveal conspicuous structures.
The distribution of dusty red galaxies follows a similar two group-structure,
possibly with a slight offset to lower redshifts in the North group. The blue
galaxies mostly avoid the concentrated areas which are heavily populated by
red galaxies of both kinds, most notably the Northern group, and almost seem 
to live in the gaps left by the $(y,z)$-structure of red galaxies. 

This is an intriguing piece of evidence when combined with the $(x,y)$-maps,
where old red galaxies are clearly clustered, while the dusty red and blue 
galaxies have low clustering in common. But in the $(y,z)$-maps dusty red 
galaxies occupy a similar region as old red galaxies, while blue galaxies
tend to avoid them. Most of the redshift signal is likely to originate from 
peculiar velocities, since a line-of-sight distance equal to the lateral 
size of our field causes only 350~km/sec of redshift. Hence, a tentative
consistent explanation of the maps could place the blue galaxies in a large 
low-density volume where large distances from massive cluster cores leave 
them with low random velocities and a natural spread in positional redshift. 
The old red galaxies form concentrated cluster cores that move at relative 
bulk velocities of $\pm 700$~km/sec (N$+$/S$-$) through the blue envelope. 
The dusty red galaxies would be associated in velocity and position with the 
cores populating their outskirts with less concentration.

Finally, we determine velocity histograms for each galaxy type independently 
(bottom row in Fig.~\ref{xymaps}). The full samples suggest high velocity
dispersion, where especially the dusty red galaxies appear to have $\sigma_{cz}
/(1+z) \approx 1200$~km/sec. There are indications of a bimodal distribution
among both types of red galaxies, arising from the two groups in velocity space. 
When analyzing only the encircled groups, we find $\sigma_{cz}/(1+z) \sim 550
$~km/sec (rest-frame) for all four red samples. Such a split does not make 
sense for blue galaxies that show no structure in velocity space anyway. 

However, there are differences between the North and South group: all three
galaxy types in the South group have mean redshifts and velocity dispersions 
that are consistent with identical given our number statistics. In the North
group however, we find almost no bright blue galaxies (hardly any member in
the spectroscopic sample) and dusty red galaxies could be displaced to the
North and to lower redshift compared to the old red galaxies. We note, that 
the empty regions in these z-diagrams are truely devoid of galaxies and not 
artificially empty due to selection effects. The only selection at work was 
a luminosity cut for the spectroscopic sample.

A complete analysis of the structure of this supercluster complex is beyond
the scope of this paper. But we have learned, that the spatial and velocity
distribution of the three galaxy types is consistent with dusty red galaxies
forming an intermediate population whose preferred habitat are the regions
where the high-density cluster cores interface with the lower-density field.
The preferred habitats of old red and young blue galaxies are to either side 
of the interfacing dusty red galaxies.

\begin{figure}
\centering
\includegraphics[clip,angle=270,width=\hsize]{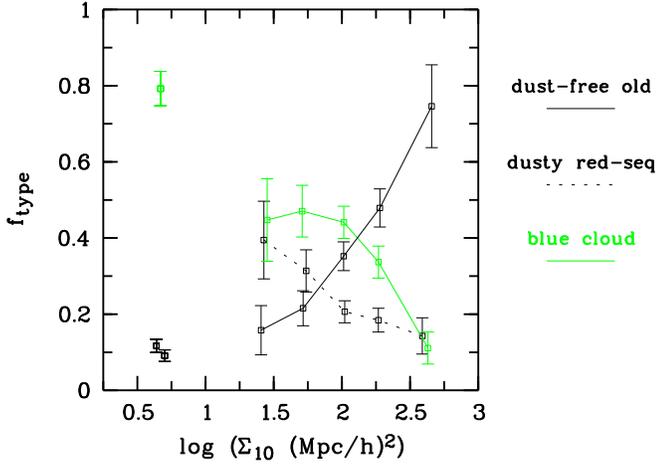}
\caption{Type-density relation: The cluster environment covers projected 
galaxy densities of $\log (\Sigma_{10} ({\rm Mpc}/h)^2) = [1.25,3.05]$.
The field points lie at a density of 0.67, but old red (9\%) and dusty red 
(11\%) galaxies are slightly offset for clarity.
The fraction of intermediate-age dust-reddened galaxies in the red-sequence
increases from low-density fields to the outskirts of a cluster environment, 
but then decreases gradually with density inside the cluster. 
\label{typedensity}}
\end{figure}

\subsection{Type-density relation}

We now investigate how the fraction of each type among the galaxy population
depends on environment, assessed by the projected galaxy density $\Sigma_{10}$.
We define $\Sigma_{10}$ for any galaxy as the number of galaxies per $\sq
\arcmin$ or (Mpc$/h)^2$, within a circle whose radius is the 
average of the distance to the 9th and 10th nearest neighbor. In
Fig.~\ref{typedensity} these fractions are plotted in five bins of galaxy 
density, which have bin limits of $\Sigma_{10} \sq\arcmin=[0.25,0.5,1,2,4,16]$. 
The geometric average of the whole cluster sample is $\langle \log (\Sigma_{10} 
\sq\arcmin)\rangle =0.222$ or $\langle\log (\Sigma_{10} ({\rm Mpc}/h)^2)\rangle
= 2.07$. In addition, we plotted the type fractions for a field sample and place 
the data points at $\log (\Sigma_{10} ({\rm Mpc}/h)^2)=0.67$, the mean density 
for the estimated level of field contamination. We remind the reader, that we 
have not applied any statistical subtraction of the field contamination.

Already from the cluster data alone it is clear that the fraction of dust-free 
old populations increases with galaxy density, while that of the blue cloud
galaxies decreases. The dust-reddened intermediate-age galaxies within the 
red-sequence appear to decline more gently towards high density than the blue 
cloud galaxies do. However, including the field fractions, we realize that the
trend of dust-reddened red-sequence galaxies is more complex than the simple
picture of old galaxies increasing and young galaxies decreasing with density.
Here, we find that these dust-reddened contaminants increase strongly when
going from the very low-density field environment to the medium-density region 
at the outskirts of the cluster, but then roll off gently towards the higher 
densities again. The low
density gradient in their fraction renders them as an intermediate population 
between the blue cloud and the dust-free old clump: are these galaxies which 
experience a transformation from typical field to cluster properties? 

\begin{figure*}
\centering
\includegraphics[clip,angle=270,width=0.8\hsize]{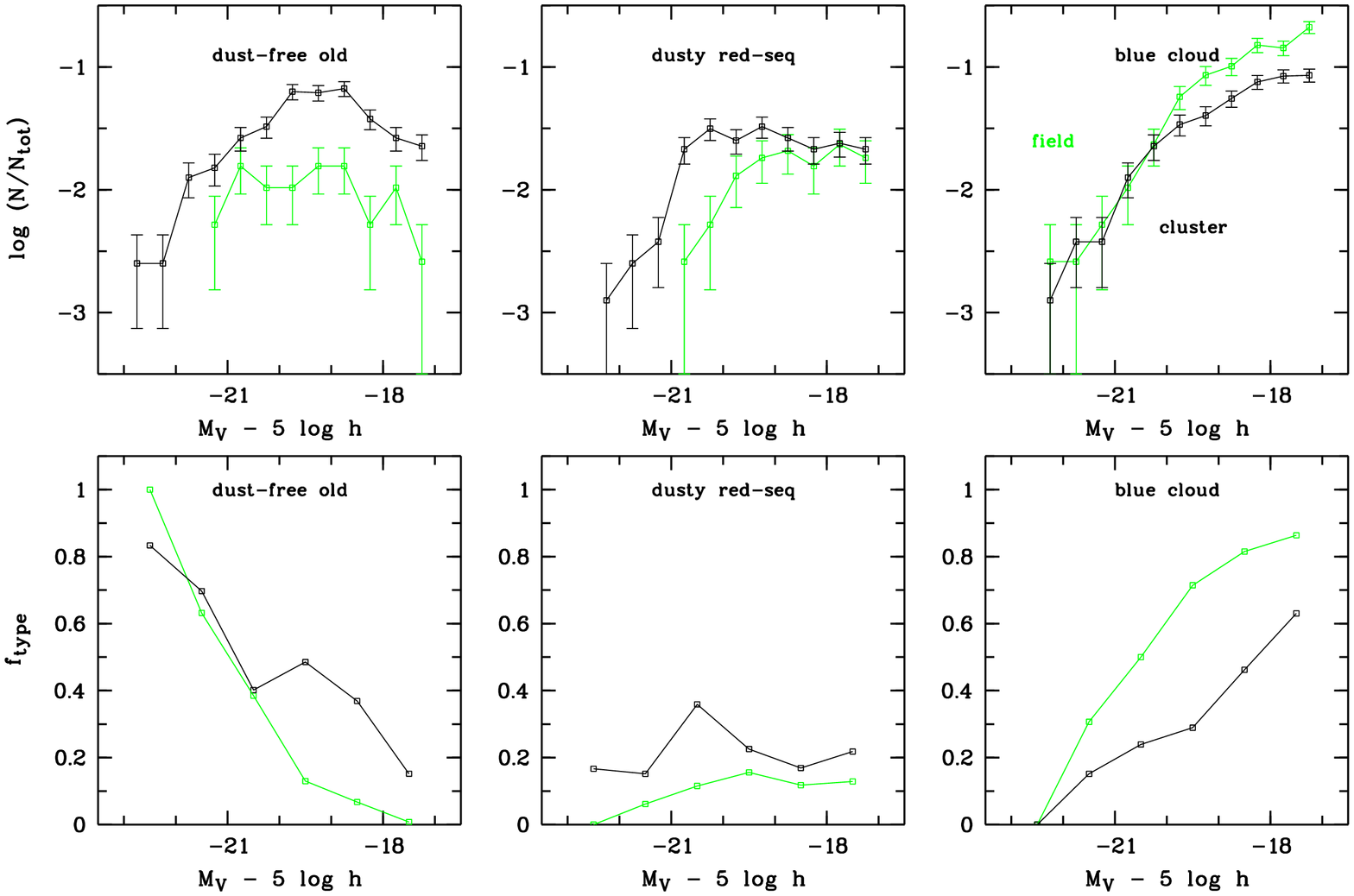}
\caption{{\it Top row:} Luminosity distributions: 
The dust-free old populations follow almost a Gaussian, while the blue cloud 
could more easily be described by a Schechter function. The cluster shows an
excess of dust-free old populations, especially fainter than L*, and of dusty
red-sequence contaminants, mostly brighter than L*. The field and cluster 
samples are each normalised by the number of their galaxies.
\label{LFs}}
\end{figure*}

\subsection{Luminosity distributions}

We determine luminosity distributions of cluster and field galaxies by type, 
where both have been normalised to their respective total number of galaxies 
at $M_V<-17$. The cluster is not corrected for incompleteness induced by 
increasing redshift errors towards the faint end, which may have shifted 
points by -0.15~dex at $M_V\approx -17$ if we assume $\sim$68\% completeness 
(see Sect.~\ref{contamination}). The field sample is $>90$\% complete at all 
luminosities shown here, and redshift errors do not affect its completeness 
given a smooth underlying field. No attempt is made to correct these observed
luminosities to any sort of bolometric luminosities, e.g. by taking into 
account the dust extinction. Since completely obscured stellar populations
would neither affect the measured luminosities nor the reddening observed
in the visible stellar light, we leave any improvement involving corrections
for dust until we have MIR data available.

Looking at differences between the types, we find a deficit of faint blue cloud 
galaxies in the cluster. The faint-end slope is not 
significantly different after adjusting the cluster for a roughly estimated incompleteness. This deficit is compensated by an enhancement of red-sequence 
galaxies at all luminosities. We find an excess of dust-free old populations at
all luminosities but especially among sub-L* galaxies fainter than $M_V=-20$.
We also find an excess of dusty red-sequence contaminants which is especially
strong among galaxies brighter than $M_V=-19$. At the faint end of our sample
dusty red-sequence galaxies appear to have similar fractions in the cluster
and the field sample. 

The difference between the cluster and the field can be summarized with two 
phenomenological aspects: First, in the outskirts of the cluster a fraction 
of the more luminous blue cloud galaxies appear as dusty intermediate-age
contaminants to the red-sequence. Here, the stellar population could be of
genuinely intermediate age or it could be composed of a large old population
with a smaller younger population. Secondly, 
in the more concentrated cluster regions a fraction of less luminous blue 
galaxies appear instead as red galaxies and part of the dust-free old clump, 
indicating that their transformation has already happened longer ago.

\section{Discussion}

\subsection{A third party: dusty red-sequence galaxies}

Recently, galaxies have been split into a red, old passively evolving and
a blue, young, star-forming population. The long-established colour-density
relation means red galaxies are strongly concentrated in dense cluster cores
while blue galaxies overwhelmingly dominate the low-density field. In this
paper we have added complexity to the simple colour-density relation from a
{\it third type} of galaxies identified here to be related to clusters.
They do not fit into the simple bimodal description of red-vs.-blue galaxies, 
and instead show signs of an ongoing transition induced by a cluster-specific 
mechanism. This process involves the appearance as a red-sequence galaxy which 
derives its colour from both dust reddening and 
old age of the stellar population, while actively forming stars.
These properties are indicated by galaxy colours, 4000\AA -break and O{\sc ii}
emission. Galaxies of this kind may not be restricted to the red-sequence, 
but in A901/902 they constitute more than a third of the red-sequence 
population while only $<2/3$ are dust-free and old.

These dusty star-forming contaminants to the red-sequence have very little in 
common with the dominant red-sequence population. The dusty galaxies share a
spatial distribution and an avoidance of dense cluster regions with the blue 
cluster galaxies. But in contrast to blue galaxies, they are not common in 
the low-density galaxy field outside of the cluster. We may think of them as 
an intermediate population which could form a spatial interface between blue 
field galaxies and red cluster galaxies in cores.

There is further evidence that the dusty red galaxies form an intermediate 
population between blue and red galaxies. They contain a significant old 
stellar population as indicated by their Ca H\&K lines and 4000\AA -break and 
they are mostly of medium luminosity, at least in Abell 901/902. We repeat 
here, that their mean level of dust extinction is measured to be modest 
($E_{B-V}\sim 0.2$) and very similar to that in typical blue cloud galaxies.

It has been known for a long time, that a small fraction of red-sequence 
galaxies are edge-on spiral galaxies, which are red because we only have a 
clear view of their old red bulge, while their younger disk is reddened by 
dust lanes. So, edge-on spirals form a fraction of the dusty red galaxies 
in both the field and the cluster sample. However, any orientation effect 
can not lead to environmental differences in the fraction of edge-on spirals 
versus classical blue spirals. Hence, the dominant component of dusty red
galaxies must be the result of an evolutionary phenomenon. Where did they 
come from and where do they go? Generically, we consider two possible origins:

\begin{enumerate}
\item 
either they originate from the blue cloud, their transformation was induced 
by some process when they got to the outskirts of the cluster, where they are 
now caught in the act of turning into typical old red cluster galaxies, or
\item
they are minor mergers of established old cluster galaxies with infalling 
smaller blue field galaxies leading immediately to intermediate properties
without invoking any additional process.
\end{enumerate}

We will not settle any decision between these two alternatives in this paper, 
and we may well be looking at a mixed bag of objects from both origins plus
edge-on disks. Furthermore, it is not clear that all transforming galaxies
will appear unusual. If star formation just fades away as in a suffocation
scenario, the galaxy may well evolve straight onto the red-sequence without
any detour through a dusty, star-forming stage. We would need to know how the
timescale for clearing the dust compares with the vanishing of the young 
stellar population to clarify this. However, we list some arguments supporting
either one of the two scenarios:

\begin{itemize}
\item  {\it Pro (1):}
somehow galaxies in clusters have taken the turn from being blue and actively 
star-forming with related dust in the past to being old, red and dust-free now. 
Surely we should be able to catch them in the act somewhere, maybe this is it.
\item  {\it Pro (1)/Contra (2):}
there are only few old red galaxies found in the medium-density habitat 
preferred by the dusty red galaxies. It seems unlikely that most red galaxies 
are simultaneously engaged in mergers. Fig.~\ref{typedensity} suggests
that across the medium-density regime the fraction of red vs. blue galaxies 
remains roughly constant while the fraction of dusty red galaxies drops off 
towards higher density, maybe due to increased velocities deeper in the 
potential well making mergers harder.
\item  {\it Pro (2)/Contra (1):}
the most luminous dusty red examples reach higher luminosities than the blue
cloud field galaxies, which requires a cluster-specific high-luminosity 
component among the progenitors.
\item  {\it Pro (2)/Contra (1):}
the stellar ages suggested by the template SEDs and the spectral properties 
show a larger proportion of old stars than expected for a typical field blue 
cloud galaxy which is just undergoing a first transition.
\end{itemize}

Where do they go? Most likely the dusty red galaxies will look like typical
old red cluster members when their younger stars have vanished and the dust 
has cleared. They would not turn into typical blue galaxies again, because 
they already contain a significant old stellar population. 

Given the abundance of galaxies of the third kind, the two possible origins 
translate furthermore into two routes of cluster assembly. Route (1) means, 
that clusters (defined by a concentration of massive old red galaxies) grow 
via transformation of infalling galaxies intro proper cluster galaxies, e.g.
major mergers leading to dusty starbursts among infalling blue cloud galaxies.
Route (2) means that clusters grow via the growth of its individual galaxies,
where infalling galaxies are incorporated into existing large and red cluster 
galaxies via minor mergers.

We will try to collect further clues from the literature in the following.
Kodama et al. (2001) have found that the dominance of blue galaxies changes
into one of red galaxies at densities typical for galaxy groups rather than 
in cluster cores. They quote a transition density of $\Sigma_{10} \approx 
10^{1.5}$~Mpc$^{-2}$ for $M_V<M_V^*+4$, $H_0=50$~km/(s Mpc) and $q_0=0.1$ 
(taking into account their Erratum on the calculation of $\Sigma$). In our
cosmology this equates into $\Sigma_{10} \approx 10^{2.03} h^2$~Mpc$^{-2}$
for $M_V \la -16.8$. We can study lower overdensities in our field thanks 
to the much reduced field contamination given much more accurate photo-z's. 
We find our dusty red galaxies at densities below the transition quoted by 
Kodama et al. (2001), while indeed the fration of old red galaxies increases
closer to their transition density. 

While the densities where the change happens may be typical for groups, we
do not see particular group signatures among the dusty red galaxies in A901.
If the dusty red galaxies just formed an interface between the blue and red
habitats, this would support origin (2) which requires combining progenitors 
from both habitats. In contrast, should a more careful analysis find our 
dusty red galaxies to live in groups escaping our attention here, this would
support origin (1) as an internal process among the infalling blue galaxies 
which is only triggered by the cluster presence. Again, it remains unclear 
whether the dusty red galaxies in A901 mark {\it one} or {\it the} process 
turning blue galaxies into red galaxies. They may not tell us anything about 
other processes happening in groups.

\subsection{Alternative origins of the red colours}

We try to consider alternative scenarios where the colours of our third party
of allegedly dusty red-sequence galaxies is not a result of dust. We consider
mixed stellar populations and additional nuclear light sources:

{\it Post-starburst galaxies of k+a type:} Such galaxies are roughly characterized 
by a linear combination of an E-galaxy SED with an A-star SED. A stars are bluer 
than old red galaxies throughout the visual wavelength range. As a result, any 
sum of these components is bluer than a pure old red galaxy. Also, Balogh et al. 
(2005) have excluded the possibility that the colours of k+a's are significantly 
affected by dust. Hence, they can not show the very red colour observed at the 
long-wavelength end in Fig.~\ref{templat} which is also seen in the $R-I$ colours 
in Fig.~\ref{clump_sel}. We expect k+a galaxies to reside on the bluer end of the 
dust-free old clump and to some extent in the blue cloud.

{\it Additional nuclear light sources:} When considering these or indeed any 
internal colour gradients, we need to be aware of possible aperture bias. Our 
SEDs and their colours are determined from aperture 
photometry which probes a region of fixed physical size for any galaxy at the 
cluster distance. As a result, all nuclear components in the spectrum will appear 
enhanced as compared to SEDs from total object photometry. HST images will soon
permit to quantify such effects. The origin of the nuclear component could be in 
dusty star-forming regions or potentially in active nuclei, and is constrained by 
our observations of continuum colours and emission lines. Active nuclei are almost 
certainly ruled out, because (i) our spectra show only very weak Oxygen lines, 
(ii) the objects are not detected in a 90~ksec XMM image, and (iii) the abundance 
of these objects. Furthermore, the continuum shape does not permit AGN light as a 
sole modification on top of an old red SED: The deviation from an old red SED is 
increasingly blue on the blue side and increasingly red on the red side.

\subsection{Previous observations of dusty star-forming cluster galaxies}

Coia et al. (2005) have discussed the properties and abundance of mid-IR 
sources in clusters from ISOCAM observations. Generally, MIR-luminosity is 
a good tracer of total star formation, given that the largest part of star
formation is usually dust-enshrouded in larger galaxies. They found that 
half of their sources (heavily star-forming cluster galaxies) reside in 
the red-sequence. While they are obviously not dust-poor old populations
following passive evolution, their red colour is instead explained by dust
reddening. In contrast to the main red cluster population, the dusty red 
starbursts were not concentrated in high-density regions.

If such galaxies existed in Abell 901/902, we would identify at least the
red-sequence part of the population. Most likely they are similar objects 
as our dusty star-forming red-sequence contaminants: a property common to
both is their tendency to avoid the high-density cluster cores. In the
cluster Cl 0024+1654 the MIR sources have a broader velocity dispersion 
than the main cluster population, just like the dusty red galaxies do in 
the Abell 901/902 complex.
We note, that Coia et al. (2005) found red starburst galaxies associated 
with close neighbors in all four cases where HST imaging was available.

Poggianti \& Wu (2000) have discussed in detail a type of galaxy identified
initially by Dressler et al. (1999) in distant cluster studies, which may 
resemble our dusty red galaxies to some extent. They have found {\it e(a) 
galaxies} showing both O{\sc ii} emission and H$\delta$-absorption. Using 
line ratios and FIR fluxes, they explained these galaxies consistently as 
dusty starbursts with an age-dependent extinction level. Furthermore, they
found 75\% of their e(a) galaxies to be mergers or interacting pairs with 
close companions, supporting a merger origin to the starbursts.

Our dusty red galaxies show similar features, but their mean H$\delta$ 
absorption of only $EW\sim 2.6$~\AA \ does not let them pass the e(a) 
criteria. Rather than speculating about the precise nature of our dusty 
star-forming red-sequence galaxies we defer this discussion until 
observations with HST and {\it Spitzer} have been analyzed, which will deliver
a wealth of detailed information about total star formation rates and
morphologies.

\section{Summary}

We have analyzed the galaxy population of the supercluster system Abell 901a,
901b and 902 located at redshift $\sim 0.17$. From the COMBO-17 catalogue, we 
have selected the bright galaxy population with $M_V<-17$ by using a photo-z
selection of $z=[0.155,0.185]$. Given the photo-z accuracy of $<0.01$, this
sample is virtually complete at $M_V<-19$ and could be $\sim 30$\% incomplete
at $M_V=-17$. For the $\sim 250$ most luminous cluster galaxies we obtained 
additional spectroscopic data in the blue part of the spectrum using 2dF. The
supercluster is a system of great morphological complexity with several large 
concentrations which are presumably unrelaxed and will merge in the future. 
Our results are as follows:

\begin{enumerate}
\item
We improve the selection of old passively evolving galaxies by using details
of the COMBO-17 SED information. We find, that the red-sequence contains both 
galaxies which are red because they are old and passive, and those which are 
modestly dust-reddened (168 objects, more than 1/3 of the entire red-sequence).
\item
We find the long-established colour-density relation again in this system (we
did not investigate the morphology-density relation, Dressler 1980). However, 
the overall increase of the red galaxy fraction has two contributions, which 
add complexity to the colour-density relation:
(a) an increase of old passively evolving galaxies in high-density cores, and
(b) an excess of dusty star-forming galaxies in the medium-density outskirts.
\item
Thus, we have identified a part of the galaxy population which does not follow
the canonical colour-density relation. A rearticulation of the colour-density 
relation with three classes could be, that young blue galaxies prefer low-density 
environments, dusty red intermediate-age galaxies prefer medium-density 
environments and old red galaxies prefer high-density environments. 
\item
The fraction of dusty red-sequence galaxies among the total is much higher in 
the cluster ($\sim 22$\% at $M_V<-17$) than in the field ($\sim 11$\%). Within 
the cluster, their fraction declines with increasing density just as the blue 
galaxies do. In their preferred habitat of medium-density regions they match 
the fraction of blue galaxies ($\sim 40$\%).
\item
The dusty red cluster galaxies have as low spatial clustering as the blue 
cluster members do. However, using position-velocity-diagrams we find that 
dusty red galaxies tend to be associated with the cluster cores formed by 
old red galaxies, while blue galaxies prefer the voids.
\item
After accounting for the bimodal structure in redshift space, the velocity 
dispersions of dusty red and old red galaxies come out similar with $\sim 
550$~km/sec for each velocity group. Looking at the supercluster as a whole, 
dusty red galaxies have the largest dispersion and preferably populate 
the extreme velocity ends.
\item
Spectra of the dusty red-sequence galaxies show they are currently forming
stars but they have plenty of old stars as well. They show mean stellar ages 
which are intermediate to old red and young blue galaxies. It is not clear,
whether this means that (i) they are a mixture of very old stars and a much
younger population, or (ii) their stars are of truely intermediate age while 
young stars are missing or hidden by dust.
\item
The dusty red galaxies either originate from minor mergers of established 
old galaxies with smaller infalling blue galaxies, or result from internal 
transformations of infalling galaxies which is triggered by the cluster
environment. They might indicate where either {\it some} or {\it the} 
process attacks, which transforms galaxies from blue to red.
\end{enumerate}

Further observations of the Abell 901/902 field are pending. A mosaic with 
HST/ACS (PI Gray) will provide a wealth of morphological information and 
clues to the transformational mechanisms at work in the dusty red galaxies. 
Observations with the {\it Spitzer} will reveal hidden star 
formation and clarify the total star formation activity in all of the
cluster galaxies. Preliminary results indicate that more than half of our 
red-sequence galaxies with $E_{B-V}>0.25$ and $M/M_\odot >10^{10}$ are 
detected in {\it Spitzer} 24$\mu$ images (Bell et al., in preparation).

This work has only used ground-based optical data. We used the discriminative
power and concentrated information content of the medium-band survey COMBO-17
to study the cluster population with high completeness and low contamination.
Just from the filter SEDs we could select cluster members and identify dusty
contaminants to the red-sequence. We have thus established a third class of
cluster members beyond the simple red-blue division. From previous work of the
colour-density relation it would have been counter-intuitive to predict a rich 
population of dusty red galaxies in higher-density environments. However, we 
have shown them to be a phenomenon which is genuinely related to galaxy clusters
and preferentially populate their medium-density envelopes interfacing between
the young blue field and the old red cores.

In the near future, many medium-band surveys which are wider or deeper than
COMBO-17 will target further galaxy clusters, such as a wide-area medium-band 
survey with the VST. These will enlarge
the sample of clusters studied with the technique presented here, and will
reveal possible variations in the dusty red galaxy content, which would provide
further clues to the physical origin of this phenomenon.

\begin{acknowledgements}
C. W. was supported by a PPARC Advanced Fellowship. M.E.G. was supported by 
a PPARC Postdoctoral Fellowship and an Anne McLaren Research Fellowship.
We appreciate discussions with Alfonso Aragon-Salamanca, Michael L. Balogh,
Steven Bamford, Eric F. Bell, Mustapha Mouhcine and Stella Seitz, and thank
an anonymous referee for useful comments.
This publication makes use of data products from the Two Micron All Sky 
Survey, which is a joint project of the University of Massachusetts and the 
Infrared Processing and Analysis Center/California Institute of Technology, 
funded by the National Aeronautics and Space Administration and the National 
Science Foundation.
\end{acknowledgements}

\appendix

\section{Accuracy of photometric redshifts}

\begin{figure*}
\centering
\includegraphics[clip,angle=270,width=0.75\hsize]{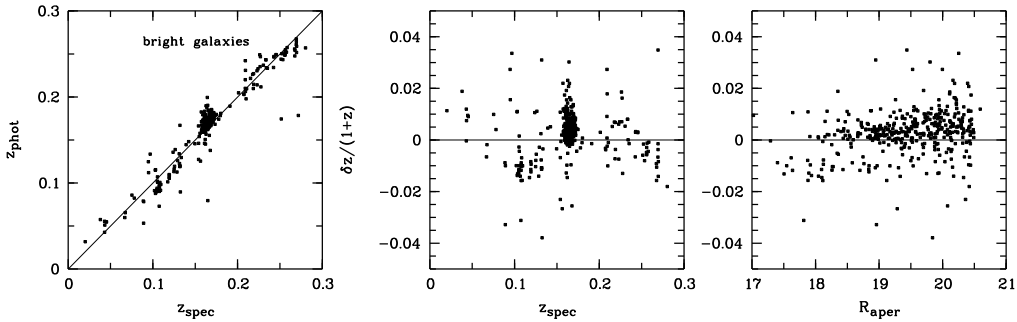}
\caption{Redshift quality: 404 bright galaxies observed with 2dF on three
COMBO-17 fields, mostly containing galaxies from Abell 901/902.
{\it Left panel:} Comparison of $z_{phot}$ vs. $z_{spec}$. 
{\it Center panel:} Redshift error $\delta z/(1+z)$ vs. redshift $z_{spec}$.
{\it Right panel:} Redshift error $\delta z/(1+z)$ vs. $R$-band aperture
magnitude. The 1-$\sigma$ redshift error is $<0.01$, and the outlier rate
is $<$1\% (3 out of 404 objects with errors around 0.06).
\label{p3_gals}}
\end{figure*}

Spectroscopy of bright galaxies was available through the 2dFGRS survey on 
the S11 field with 39 galaxies including the cluster A1364 at $z\sim 0.11$ 
and 14 galaxies on the CDFS field, most at $R\la 18$ \citep{Col01}. More 
observations on the A901 field provided a larger and deeper sample of 351 
galaxies at $R\la 20$ including the cluster A901/902 at $z\sim 0.165$. The 
total number of independent redshifts is 404. Fig.~\ref{p3_gals} displays 
the redshift quality of these bright galaxies. We find that 77\% of the 
galaxies have redshift errors $\delta_z/(1+z)$ below 0.01 while three 
objects (i.e. less than 1\%) deviate by more than 5-$\sigma$ from the true 
redshift.

Looking specifically at the A901/2 cluster system, we notice a consistent
small offset of $\langle \Delta_z \rangle =0.005$ between photometric and 
spectroscopic redshifts. The reason for such offsets lies in systematic
differences between the predicted and measured photometry of the galaxies.
Measurements can be off due to small photometric calibration errors, and
colour predictions can be off due to slightly unrealistic template SEDs or
even slightly wrong filter transmission curves. At $\lambda = 600$~nm, the
offset corresponds to a 2.5~nm error in estimating the location of spectral
features from the multi-band photometry. However, another galaxy cluster,
Abell 1384 lies partially within the COMBO-17 {\it S11 field} at $z\sim 
0.105$ and is estimated with a larger offset of $\langle\Delta_z \rangle 
=-0.01$ in the opposite direction, suggesting that field-dependent errors 
in the calibration are more relevant than filter curve errors.

Spectroscopy of fainter galaxies is available via the VIMOS VLT Deep 
Survey \cite[VVDS,][]{LeF04} on the CDFS field. Using galaxies with VVDS 
redshift reliability of 95\% or greater, we find five outliers among 334 
galaxies (1.5\%) at $R<23$ with true redshift errors on the order of 0.1, 
while the remaining 98\% have a distribution of photo-z errors $\delta_z 
= z_{\rm spec}-z_{\rm phot}$ with a mean offset of $\delta_z/(1+z) = 
-0.006$ and an rms scatter of $\sigma_z/(1+z)=0.020$.

The redshift grid for the galaxy colour library covers the range from
$z=0$ to $z=1.40$ in 177 steps. These are equidistant on a $\log (1+z)$
scale with steps of 0.005 and of course limit the redshift resolution 
when reconstructing galaxy density features in redshift space. According 
to the sampling theorem, features in redshift space can be recovered if
their wavelength is at least twice as large as a grid step. Thus, we have
to expect that features with wavelength $\delta_z/(1+z) < 0.01$ will be
smoothed by our redshift estimation even under perfect conditions where
systematic problems in photometric calibration or SED match are absent.
This will not significantly restrict the power of our dataset, 
because our redshift estimation is probably not consistently much more 
accurate than 0.01 across all redshifts and SEDs we are interested in.

\section{In retrospect: selecting old stellar populations} 

\begin{figure}
\centering
\includegraphics[clip,angle=270,width=\hsize]{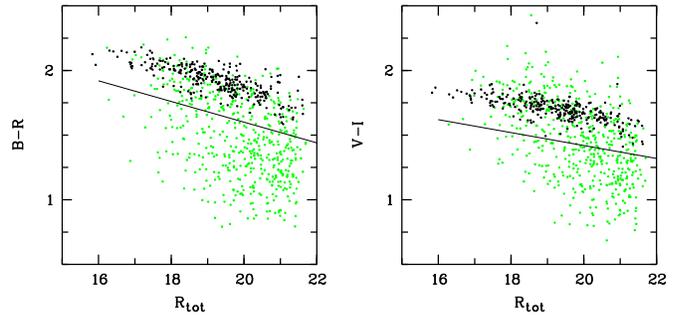}
\caption{Observed-frame colour-magnitude diagrams at $z\sim 0.17$: 
{\it Left:} An observed-frame colour-index straddling the 4000\AA -break 
makes dust-free old populations (black dots) appear as red as possible in 
comparison to dust-reddened galaxies. A Butcher-Oemler-style red-sequence 
cut (line) $0\fm25$ below the colour-magnitude relation selects some younger 
dustier galaxies, which can be identified with the photometric information 
in COMBO-17.
{\it Right:} A colour index longwards of the 4000\AA -break moves the 
red-sequence bluewards, further into the bulk of the galaxy population,
making an efficient selection of dust-free old populations very challenging.
\label{RS_dfo}}
\end{figure}

Galaxies with passively evolving old stellar populations have traditionally
been selected from colour-magnitude diagrams. In the absence of any prior
redshift information, observed-frame colours allow to search for clusters of 
galaxies using a red-sequence selection \citep[e.g.][]{GY00,GY05}. This method
capitalizes on the fact, that usually galaxies in any cluster red-sequence
show redder observed-frame colours than foreground galaxies. If redshifts and
rest-frame data are available, the cluster population can be separated from
the field with vastly improved clarity and can include blue cluster members
as well.

In rest-frame colour-magnitude diagrams (at least anywhere at $z<1$) the galaxy 
population appears bimodal with a narrow red-sequence and a wide blue cloud 
\citep{Str01,Bal04,Bell04}, provided a suitable colour index is used. The 
selection of an
old stellar population via a red-sequence cut works best when a colour index
is used that depends most sensitively on mean stellar age, i.e. a colour index 
made from two passbands enclosing the 4000\AA -break, such as rest-frame $U-V$. 
Colour indices made from two passbands which reside both on one side of this 
break, such as rest-frame $R-I$ or NUV colour indices depend to a lesser degree 
on age and to an increased degree on dust reddening.

We illustrate this conclusion with colour-magnitude diagrams of A901,
where the location of the dust-free old clump is highlighted. Here, we would 
like to suppress possible errors introduced by translating measured data into 
rest-frame quantities and plot measured magnitudes and colours instead. At $z
\sim 0.16$, the observed-frame $B-R$ corresponds approximately to rest-frame
$U-V$ and $R$ to $M_V$ (except for a zeropoint shift). Enclosing the 4000\AA
-break we observe the strongest possible sensitivity to age and find the red 
sequence at the red end of the galaxy distribution (see Fig.~\ref{RS_dfo}). 
However, a Butcher-Oemler-style red-sequence cut defined by a line parallel to
the colour-magnitude relation, but $0\fm25$ bluer, still selects a number of 
younger, dust-reddened systems in our cluster sample.

If we use the colour index $V-I$ instead, which corresponds approximately to
$(B-R)_{\rm rest}$ the relative impact of dust reddening on the colour is 
increased compared to that of stellar age. As a result, the red sequence moves
bluewards, further into the bulk of the galaxy population. A Butcher-Oemler
style selection will now be contaminated even more by dust-reddened galaxies.
However, using more detailed photometric information as in this paper helps to
achieve a much cleaner selection of dust-free old galaxies.

\end{document}